\documentclass[10pt,journal,compsoc]{IEEEtran}

\usepackage{hyperref}

\usepackage{balance}

\usepackage{listings} 
\usepackage{color} 
\definecolor{mygreen}{rgb}{0,0.6,0}
\definecolor{mygray}{rgb}{0.5,0.5,0.5}
\definecolor{mymauve}{rgb}{0.58,0,0.82}
 
\definecolor{darkgray}{rgb}{.4,.4,.4}
\definecolor{purple}{rgb}{0.65, 0.12, 0.82}
 
\lstdefinelanguage{JavaScript}{
keywords={typeof, new, true, false, catch, function, return, null, catch, switch, var, if, in, while, do, else, case, break},
keywordstyle=\color{black}\bfseries,
ndkeywords={class, export, boolean, throw, implements, import, this},
ndkeywordstyle=\color{darkgray}\bfseries,
identifierstyle=\color{black},
sensitive=false,
comment=[l]{//},
morecomment=[s]{/*}{*/},
commentstyle=\color{purple}\ttfamily,
stringstyle=\color{red}\ttfamily,
morestring=[b]',
morestring=[b]"
}

\IEEEoverridecommandlockouts

\usepackage{amsmath,amssymb,amsfonts}
\usepackage{algorithmic}
\usepackage{graphicx}
\usepackage{textcomp}
\usepackage{xcolor}
\usepackage{hyperref}
\usepackage{cleveref}
\usepackage{xspace}
\usepackage{paralist}
\usepackage{multirow}
\usepackage{subcaption}
\usepackage{tikz}
\usepackage{courier}
\usepackage{paralist}

\definecolor{Gray}{gray}{0.3}
\tikzstyle{mybox} = [draw=black, very thick, rectangle, rounded corners, inner ysep=5pt, inner xsep=5pt, fill=gray!20]

\newcommand{\todone} [1] {{}}

\newcommand{\bemph}[1]{{\bf #1}}

\newcommand{\link}[1]{{\color{blue} #1}}

\newcommand{\oldname}{{${\sc{ BF\!{+}\!F\!F}}$}\xspace}
\newcommand{\newname}{{\sc  SynShine}\xspace}
\newcommand{\linefix}{{\sc  LineFix}\xspace}
\newcommand{\unkfix}{{\sc  UnkFix}\xspace}
\newcommand{\blkfix}{{\sc  BlockFix}\xspace}
\newcommand{\fragfix}{{\sc  FragFix}\xspace}

\newcommand{\java}{{\sc  Java}\xspace}
\newcommand{\cp}{{\sc  C}\xspace}

\newcommand{\javac}{{{\small\tt javac}}\xspace}
\newcommand{\etal}{\emph{et al.}\xspace}

\newenvironment{smcodetabbing}
  {\relax\ttfamily\small\tabbing}
  {\endtabbing}

\ifCLASSOPTIONcompsoc
  \usepackage[nocompress]{cite}
\else
  \usepackage{cite}
\fi
\ifCLASSINFOpdf
\else
\fi
\begin{document}
%
\title{\newname: improved fixing of Syntax Errors}
%
%
%
%

\author{Toufique~Ahmed,
        Noah Rose Ledesma,
        and~Premkumar~Devanbu
\IEEEcompsocitemizethanks{\IEEEcompsocthanksitem All the authors are with the Department
of Computer Science, University of California, Davis,
CA, 95616.\protect\\
E-mail: \{tfahmed, roseledesma, ptdevanbu\}@ucdavis.edu }
\thanks{Manuscript in submission}}

%
%

\markboth{IEEE Transactions on Software Engineering}%
{Shell \MakeLowercase{\textit{et al.}}: Bare Demo of IEEEtran.cls for Computer Society Journals}
%



\IEEEtitleabstractindextext{%
\begin{abstract}

Novice programmers struggle with the complex syntax of modern
programming languages like \java, and make lot of syntax errors.  The diagnostic syntax error messages
from compilers and IDEs are sometimes useful, but often the messages are cryptic
and puzzling. Novices could
be helped, and instructors' time saved, by automated repair suggestions when dealing with syntax errors. 
 Large samples of novice errors and fixes are now available, offering the possibility of
data-driven machine-learning approaches to help novices fix syntax errors. 
Current machine-learning approaches do a reasonable job fixing syntax errors in shorter programs, but
don't work as well even for moderately longer programs. 
We introduce \newname, a machine-learning based tool that substantially improves on the state-of-the-art,
by learning to use compiler diagnostics, employing a very large neural model
that leverages unsupervised pre-training, and relying on multi-label classification rather
than autoregressive synthesis to generate the (repaired) output.  
We describe \newname's architecture
in detail, and provide a detailed evaluation. We have built \newname into
a free, open-source version of Visual Studio Code (VSCode); we make all our source code 
and models freely available. 

\end{abstract}

}

\maketitle

\IEEEdisplaynontitleabstractindextext

%
\IEEEpeerreviewmaketitle


\section{Introduction}
\IEEEPARstart{S}{yntax} errors are easy to make, and will cause compiles to fail. The
challenges posed by syntax errors to novices have been known for a long time~\cite{spohrer1986novice}. 
More recent studies have documented the challenges faced by novices
in various languages~\cite{mccracken2001multi, lahtinen2005study, jackson2005identifying}. 
Novices make a wide range of syntax mistakes~\cite{jackson2005identifying}, some of which are quite
subtle; time that might otherwise be spent on useful pedagogy  on problem-solving and logic is spent
helping novices deal with such errors. 
Unfortunately the error messages provided by compilers are often not helpful; novices
struggle to interpret the messages, and sometimes even experts do!~\cite{kummerfeld2003neglected}.
Consider for example, the real program example in Fig.~\ref{jsorig}, where a novice
student just replaced a ``*'' with an ``x'' on line 8. None of the big 4 IDEs (VSCode, IntelliJ, BlueJ, or Eclipse)
provide a direct diagnostic for this very understandable error.  
A lot of time can be spent on such errors~\cite{denny2012all}, and researchers have called
out for more attention to help novices~\cite{kummerfeld2003neglected} deal with errors,
specifically syntax errors. While \emph{semantic errors} (bug-patching) have received quite
some attention, syntax errors have attracted less interest.
\todone{Might want to acknowledge that there are two lines of work: syntactic fix and semantic fix and here you focus on syntactic fixes to basically avoid R2 from asking about bug fix work}

The possibility of collecting novice error data, and the emergence of high-capacity, highly configurable
deep learning models, has raised the possibility of designing models that can automatically fix errors, and
training them using novice data.  This approach is very attractive for several reasons: 
\begin{inparaenum}[a)]
\item Automated repair of syntax errors is helpful to novices,  and saves instructors' time. 
\item Traditional approaches to automatically finding \& fixing syntax errors require hand-coding 
fairly complex parser logic. 
\item Automatically learning models to fix errors is an approach that promises to be language-agnostic, as long
as sufficient data is available. 
\item Learning  fixing strategies from samples representative of novice errors promises to yield models that 
perform well on the most common mistakes that novices make. 
\end{inparaenum}
Several recent approaches to this problem have emerged, which are all arguably language-agnostic.

DeepFix~\cite{gupta2017deepfix} used sequence to sequence  encoder/decoder models (with roots in language translation)
to fix all sorts of errors in \cp, while 
Santos \emph{et al.}~\cite{santos2018syntax} used language models to repair just syntax errors in \java (and thus
is closer to our work). 
All of  the existing approaches take an erroneous program as input, and attempt to fix them. 
DeepFix (which uses an RNN-based Seq2Seq approach) works less well on longer programs, since
RNNs struggle with long-range dependencies. 
Santos \emph{et al.} faced similar challenges.  More recently, Ahmed \emph{et al.}~\cite{ahmed2021learning} trained a ``lenient'' parser using synthetic 
data. Ahmed \emph{et al.} use a 2 stage approach: a first  (``\blkfix'') repair
the nesting structure, and the second (``\fragfix'') to repair individual statements; we refer to their
tool in this paper as \oldname, to indicate their two stages.  Their approach
improves over both DeepFix and Santos \emph{et al.}, 
especially for longer programs.
Syntax errors in longer programs (longer than 200-300 tokens) are challenging
for automated repair, because \emph{locating} the error is difficult. All the above approaches
ignore an important source of information that could be of great value: \emph{the error from the
compiler}! 
Compiler warnings include a lot of useful information: including often the line number where
the error occurred, the tokens involved in the error, and the nature of the error. This information
could be used by neural model to better locate and repair the error. Yasunaga \etal~\cite{yasunaga2020graph} utilize \cp compiler warnings with a graph-based self-supervised approach and outperform DeepFix in fixing compiler errors. However, compiler wanrnings have not been applied to any approach that is specifically designed for \java programs. Our model also uses compiler warnings, but our performance remains robust as the programs' length increases.


In addition, existing approaches have not adequately exploited
the tremendous capacity of current DL models to learn (without direct supervision) 
the statistics of \emph{very large} amounts
of unlabeled sequential data. Modern pre-training approaches such as RoBERTa can ingest 
vast corpora of sequential data (\emph{e.g.,} a billion tokens from GitHub-hosted code)
and learn the patterns of syntax, identifier usage patterns, arithmetic expressions, 
method call patterns \emph{etc}. These patterns are automatically learned and represented as
high-dimensional vector embeddings of tokens, without requiring any human
effort to label the data. These embeddings, however, have been shown to substantially
improve performance when used as pre-set embeddings
in other networks that can be ``fine-tuned'' with smaller amounts of human-labeled
data. 

In this paper, by using the diagnostics from a compiler, and exploiting the ability
to pre-train embeddings with high capacity RoBERTa model,  we build a tool, \newname, 
which improves substantially on the state-of-the-art in automated syntax repair in \java. 
We make the following contributions: 
\begin{enumerate}
\item 
We utilize compiler diagnostics from \javac , as well as unsupervised pre-training to achieve substantial  improvements,
to implement a 3-stage syntax error repair tool, which can 
fix as much as 75\% of  programs with single errors in the Blackbox dataset.  This substantially
 improves upon prior work in the area of ``\java'' syntax error repair. 
  \item
When generating fixes, we rely on multi-label classification, rather than autoregressive synthesis, to simplify
the task of generating the repair. 
\item
We  evaluate the contributions of the different stages of our tool, and also the value of 
pre-training, and the use of \javac. 
\item
We evaluate the diversity of repairs that \newname can perform; we also dig into the cases where it appears to fail. 
\item 
We have built \newname based repair into the widely used, freely available, open-source Visual Studio Code (VSCode) tool, and made all
our software and data available to the extent allowable  under legal requirements\footnote{Blackbox data is distributed under
U.K. Laws. Please contact creators~\cite{brown2014blackbox} for details.}. 
\end{enumerate}

\noindent \textbf{Note:} Most of the novice code correction approaches are designed for \cp including DeepFix~\cite{gupta2017deepfix,gupta2018deep,yasunaga2021break,yasunaga2020graph}. Some recent works~\cite{yasunaga2021break,yasunaga2020graph}  outperform DeepFix in fixing \cp compiler errors. They all take the complete program as input and evaluate it on the DeepFix dataset with smaller sequences (up to 450 tokens).  Ahmed \etal have already shown that models taking complete program sequences tend to fail more often for longer programs~\cite{ahmed2021learning}. Unlike Blackbox, DeepFix dataset does not have erroneous and fixed program pairs. That prevents us from comparing the model's performance with the human-produced fixed versions. We train DeepFix model on our \java dataset because DeepFix uses the simplest inductive bias: sequence of program tokens and does not depend on any language-specific compiler. 
Several other
approaches~\cite{yasunaga2021break,yasunaga2020graph} are both \emph{compiler-} and \emph{language-} dependent,
so they are not comparable with our approach. Furthermore, we are able to accept complete programs, of longer length
than earlier approaches,  and provide fixes leveraging both pre-training as well as compiler errors.

\section{Background \& Motivation}
\label{motivaton}



Problem-solving, motivation \& engagement, and difficulties in learning the syntax of programming language are three fundamental challenges in introductory programming courses~\cite{medeiros2018systematic}. The dropout and failure rates are still high in introductory programming courses even after applying advanced methods and tools~\cite{watson2014failure,bennedsen2007failure}. Helping novices with programming syntax can prevent novices to get demotivated~\cite{medeiros2018systematic} at the beginning of the learning process. In this paper, we aim to help novice programmers by automatically
suggesting repairs for syntax errors. Consider the program in Fig.~\ref{jsorig}, which is an actual example
our dataset of novice programs with errors~\cite{brown2014blackbox}. Note the use of ``x'' instead of ``*'' on line 8. 
Many school maths texts use ``x'' for multiply, so this an understandable error.  

In an introductory programming course, a novice may make this error
by force of habit, and then find it quite challenging to fix the problem. 
Most popular IDEs (Eclipse, IntelliJ, Visual Studio Code) have trouble fixing this; however, our approach, which feeds a \javac-based error diagnostic, into a multi-stage repair engine that combines unsupervised pre-training, with
fine-tuning, can resolve this. 

\vspace{-0.4cm}
\begin{figure}[htb]
\captionsetup{aboveskip=-24pt,belowskip=-10pt}

\centering
\begin{lstlisting}[language=Java,basicstyle=\scriptsize\tt]
import java.util.Scanner; 
public class Multiplication
{
	public static void main(String[] args){
		Scanner sc = new Scanner(System.in);
		int a = sc.nextInt();
		int b = sc.nextInt();
		int res = a x b;
		System.out.println("The result is: " + res);
	}
}
\end{lstlisting}
\caption{Incorrect novice code sample}
\label{jsorig}
\end{figure}

Researchers have been interested in compiler diagnostics or syntax error messages for over half a century~\cite{becker2019compiler}. Barik et al. reported~\cite{barik2017developers} that the difficulties programmers face while reading or understanding error messages are comparable to the difficulty of reading source code. Understanding Java error messages is quite challenging for two reasons; i) the same error produces different diagnostics depending on the context, and ii) the compiler may produce the same diagnostic for  different errors~\cite{barik2017developers}. Though prior works~\cite{kantorowitz1986automatic,schorsch1995cap} addressed fixing errors in novice programs, DeepFix~\cite{gupta2017deepfix} was the first to apply deep learning to fix errors. DeepFix considers code repair as Neural Machine Translation (NMT) and uses an encoder-decoder based deep learning model to fix errors in \cp programs.
 Though initially aimed at semantic
bugs, the approach also works for syntax errors. 
This approach was limited by the use of RNN (recursive neural network) seq2seq models---the 
RNN architecture is challenged by longer inputs, and outputs; also since
the back-propagation through time (for the recursive elements) is not easily parallelized, 
it's challenging to exploit larger datasets and additional processors.  These became nagging problems 
in NLP; initial efforts with basic attention mechanisms~\cite{luong2015effective}
were supplanted by powerful multilayer models with multiple attention heads
to avoid recursive elements altogether~\cite{vaswani2017attention}, yielding high-capacity, 
eminently parallelizable  \emph{transformer} models. Certain errors, such
as the ones relating to block nesting, statement delimitation (with ``{\small\tt ;}'') \emph{etc.}
involve long-range syntax dependencies,  and require 
attending to very long contexts, which transformers can do better; still,
even these models fail when the dependencies become much longer.  

Ahmed \emph{et al.}~\cite{ahmed2021learning}, 
 developed \oldname, using a multi-layer, multi-head transformer approach, to  
address the limitations of traditional seq2seq models. In addition, \oldname
used a two-stage pipeline, with the first stage addressing  long-range block nesting errors, even ones
beyond the range of transformers (\emph{\blkfix}) and the second stage addressing shorter-range errors (\emph{\fragfix}).
Using the Blackbox~\cite{brown2014blackbox} dataset, they demonstrated that their approach
substantially improved over prior work on the same dataset~\cite{santos2018syntax} (which used
language models). 
\oldname  had  important limitations, noted in their paper; it didn't take advantage of
 error localization and diagnosis provided by compilers; it also didn't effectively 
 address errors in identifiers. Indeed, none of the existing approaches dealt effectively with 
identifiers, since they had to limit vocabulary. Deep learning models are challenged by
large vocabularies, which require very large embedding and softmax layers. (See~\cite{karampatsis2020big} details). We
use BPE~\cite{karampatsis2020big} to address this issue. 
 
  By addressing these limitations, we were able to achieve
 very substantial improvements on the state of the art for fixing \java programs. Ahmed \emph{et al.}\footnote{\url{https://zenodo.org/record/4420845}} and Gupta \emph{et al.}\footnote{\url{https://bitbucket.org/iiscseal/deepfix/src/master/}} provided extensive
source-available replication packages which enabled
us to provide a detailed comparison (See \S~\ref{result}).

\section{Methodology}
\label{method}
Previous work had various limitations: longer programs were
difficult to repair; error messages from compilers were not used; vocabulary limitations in DeepFix
and design choices in \oldname limited the ability to address errors in identifier usage. 
\newname directly addresses these issues, and achieves substantial improvements. 
We use a multi-stage pipeline which incorporates the Java programming language compiler (\javac), along with three learned DL neural networks (DNN). The first DNN model is
directly based on the \blkfix stage provided by  \oldname; this resolves (the potentially long-range dependent) nesting errors in the program. 
In the second stage, \newname departs from \oldname. \oldname uses the fixed nesting structure from \blkfix to split the program
into lines, and then just
tries to fix \emph{every line}; this leads to a lot of incorrect fixes.  Deepfix and Santos \etal also try to fix the entire program. 
The  second stage (\linefix) in \newname uses the line-location of the error, as detected by the standard \javac compiler, together
with the actual error message, and generates relevant fixes for delimiters, operators, and keywords; it also flags potential locations for errors
in identifier usage; these locations are sent to the third \& final stage, \unkfix. 
The \unkfix DNN model uses a Roberta-MLM to correct any
identifiers that flagged as potentially wrong by \linefix. 

\begin{figure*}[h!]
    \centering
    \includegraphics[scale=0.55]{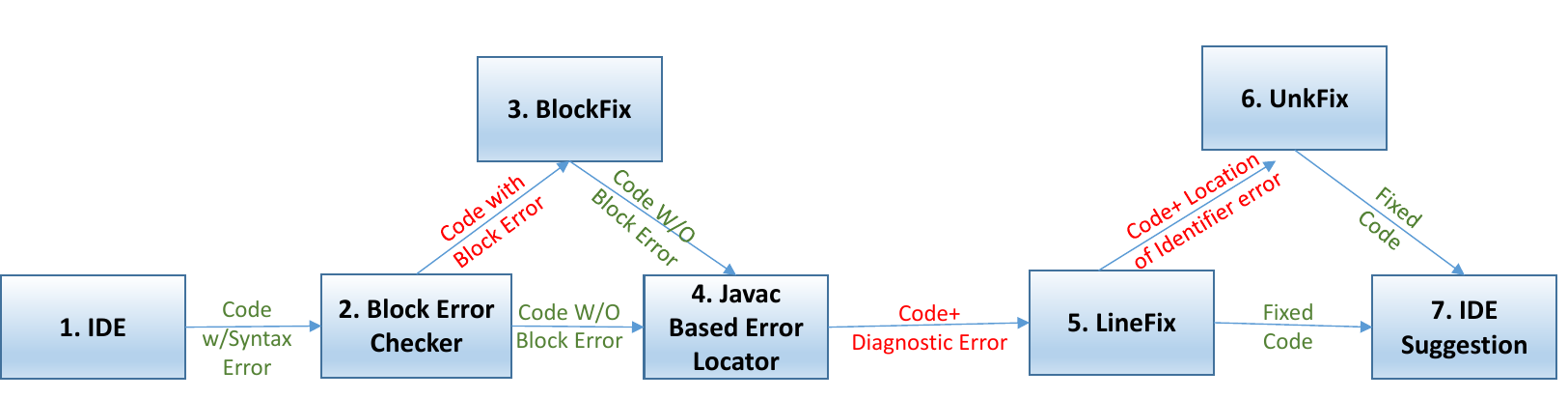}
    \caption{Overall architecture of the \newname tool.}
    
    \label{fig:pipeline}
\end{figure*}
\begin{figure*}[h!]
    \centering
    \includegraphics[scale=0.55, trim={0 9cm 4cm 0}]{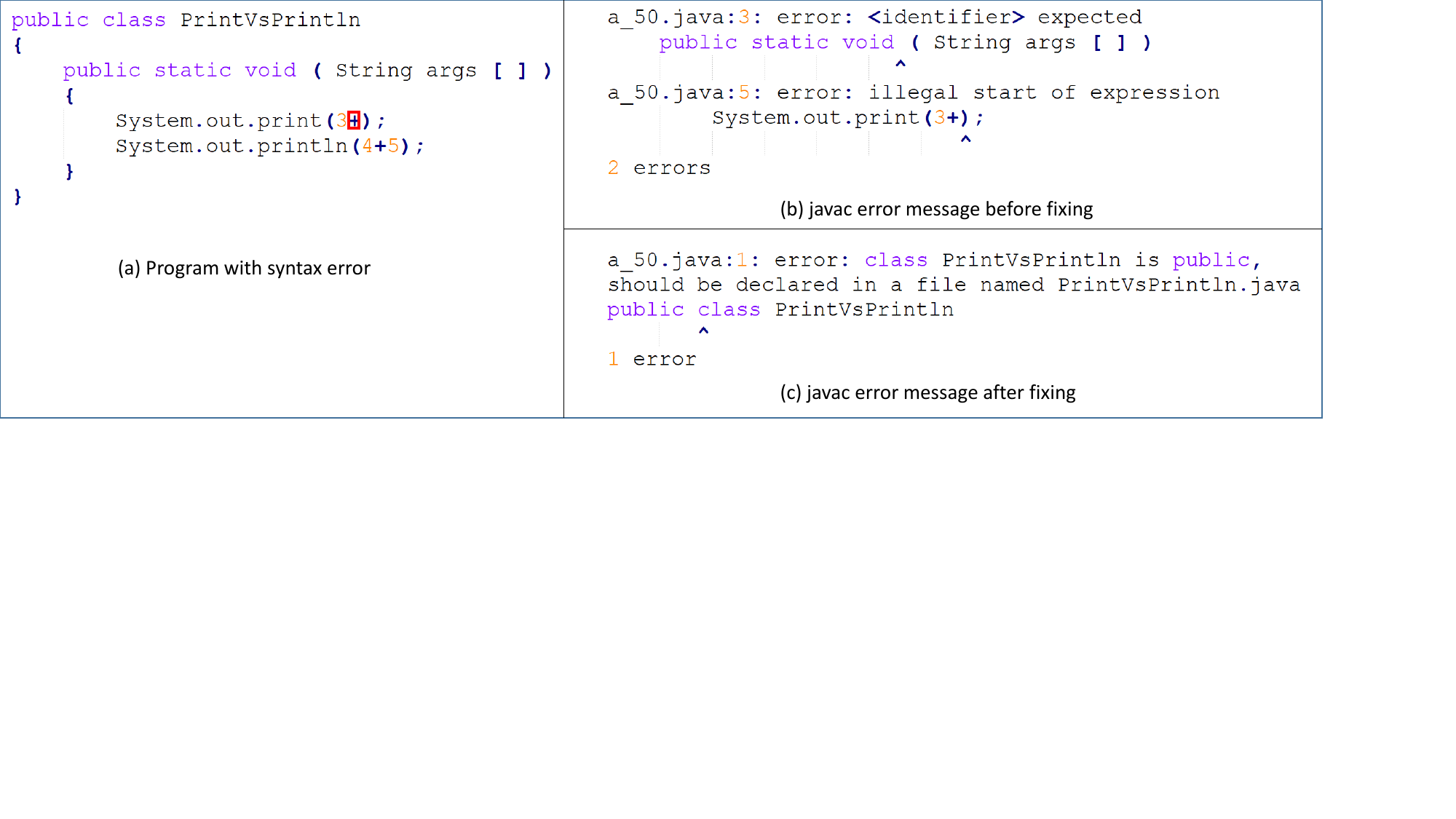}
    \caption{Locating erroneous line using \javac}
    
    \label{fig:javac}
\end{figure*}
\subsection{Overall Architecture}
\todone{why a pushdown automaton and not simply use the IDEs complaint of unclosed context?}
Fig.~\ref{fig:pipeline} shows the  architecture of our approach. When 
the IDE flags an error (step 1) we first pass the program through a block-nesting error checker (2), 
which is a simple pushdown automaton,
that checks the program's nesting structure. If  block-related issue is found, it's sent from (2) to \blkfix (3)
a transformer model (as provided in the open-source \oldname implementation~\cite{ahmed2021learning}) for repair. 
In either case, the code, hopefully now free of block-nesting errors, 
is sent to step 4, where
we try to locate the erroneous line using \javac. 
We identify the line that \javac associates with the syntax error, and pass it on to  
\linefix (step 5) with the error message. In some cases, \linefix can fix it directly; in others, it passes a token position to \unkfix (6), 
primarily to fix errors in identifier usage. Finally, the fixed code is returned as a suggestion to the IDE (7). 

We separate the line-level repairs into \linefix and \unkfix to eke out more functions out of deep-learning model capacity. 
\linefix outputs one of 154 possible editing commands,  to insert/delete/substitute delimiters, keywords, operators, or identifiers. 
We limit its 
 output vocabulary to 154.  
This limitation improves performance, but results in
 more ``unknown'' fixes, as described further below (\S~\ref{sec:linefix}).
 %
 These unknowns are resolved  by the final DNN model, \unkfix. \unkfix uses a high-capacity masked-language model 
 to suggest a fix (usually an identifier being renamed or inserted)
  given a location. In combination, these elements allow us to substantially surpass the state-of-the-art.

\subsection{{\tt javac} errors: Promises and Perils}
\label{javacnew}

While novices often find compiler error messages unhelpful~\cite{kummerfeld2003neglected}, 
our own experience suggests that they do help experienced developers! This suggests that with sufficient
training data,  machine-learning models could learn something about how to fix  syntax errors, from 
compiler syntax-error diagnostics. 
Older machine-learning-based approaches had not leveraged these diagnostics~\cite{ahmed2021learning,gupta2017deepfix,santos2018syntax}. Recently, DrRepair~\cite{yasunaga2020graph} uses these diagnostics for fixing \cp programs; \newname also uses them. 

 \javac flags syntactically incorrect programs with diagnostic errors; though the messages \emph{are not precise}, they are sometimes useful. Fig.~\ref{fig:javac} (a) presents an actual novice program with two syntactic errors (missing ``\emph{main}'' and unwanted operator ``\emph{+}''). The \javac compiler reports those two errors for the given program~\ref{fig:javac} (b). Although these error messages are unhelpful,  
\javac does in this case  finger the actual lines with errors. Line-level syntax error localization can be helpful,  if the program is long.  DeepFix, for example, 
can not fix longer programs; it relies on seq2seq translation methods,  and so has trouble with inputs longer than a few 100's of tokens. 
\oldname resolves this problem by trying to fix every line in the program using its \fragfix second stage; this approach
does induce a fair number of false positives. 
\javac promises more accurate location, which could reduce this risk. 

There is a potential issue with using  \javac,  arising mainly from the constraints of our novice error Blackbox dataset.  
\javac generates some error categories which cannot be fixed by editing the program directly. 
These errors arise for example, from file-naming conventions and incomplete typing environments. 
For example, class name \& filename mismatch errors, and missing class definition errors are shown in Fig.~\ref{fig:javac} (c). 
The Blackbox dataset (also used by Santos \etal~\cite{santos2018syntax} and Ahmed \etal~\cite{ahmed2021learning}) 
only includes programs with errors and their associated fix; it does not include the complete programming environment. 
\newname only  deals with errors that can be fixed by \emph{directly editing the \java source};  we ignore the others. 
This is a decision also made by all the other papers that deal with syntax error correction
\cite{ahmed2021learning,gupta2017deepfix,santos2018syntax}; we do, however, make use of compiler diagnostics for \java, and do manage to fix a much larger
portion of the errors in the Blackbox dataset than prior work, as seen in Table~\ref{compare1}. 
Therefore, to remove the errors we don't consider from our training set, 
we simply wrote a wrapper around \javac, to retain just those errors that can be fixed by 
editing the source \footnote{Most common ignored errors relate to ``file and class name mismatch'' and ``undeclared identifiers".}. 
%
However, it is important to note here that these ``unfixable'' errors in our dataset are counted 
in the denominator when we report our final success rate; in other words, these errors excluded from training are counted
against \newname and other tools as failures, and are not ignored in our reported performance.

\subsection{Recovering Block Structure: \blkfix}
Errors involving imbalanced curly braces are prevalent in novice programs, and are hard to resolve because of the long distance between the pair of braces. Ahmed \emph{et al.}~\cite{ahmed2021learning} report that block nesting errors consist of around 20-25\% of all syntactic errors in novice programs~\cite{ahmed2021learning}. They incorporate a component, 
\blkfix, for fixing block-nesting errors. \blkfix uses a transformer-based machine-translation model to locate \& fix block-nesting errors;  
the translation model is trained on synthetic data with artificially generated nesting errors, and the corresponding fix.   
It works with an abstracted version of the code without statements, identifiers, and types
to fix errors in nesting structure. In \newname, we simply adopt the \blkfix component from the implementation made available 
by Ahmed \emph{et al.}'s replication package.

\begin{figure}[h!]
    \centering
    \includegraphics[scale=0.28, trim={2cm 1.5cm 2cm .5cm}]{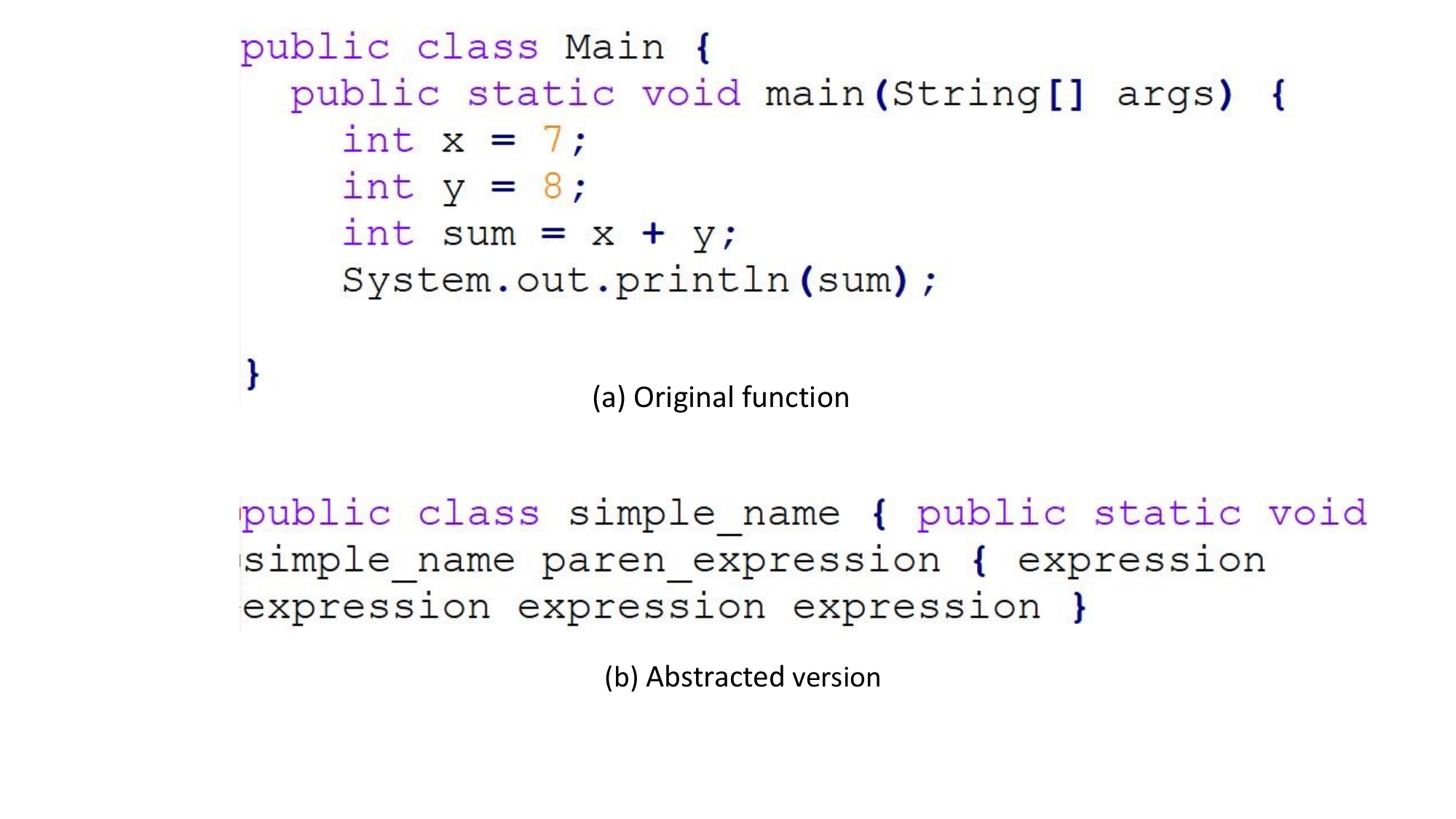}
    \caption{Abstracting source code for recovering block Structure.}
    \label{fig:blkfix}
\end{figure}

Ahmed \etal abstracted out all the identifiers, constants, expressions, and delimiters,
retraining just the curly braces and keywords (see Fig.~\ref{fig:blkfix}). They then introduce  structure-related syntax corruptions,  by adding or dropping the curly braces at randomly chosen positions; and then teaching the model to recover the original abstracted version from the corrupted model. \blkfix model learns to fix such errors by training on many such abstracted, corrupted pairs. After fixing the nesting error, the abstracted tokens are replaced with the original ones, and the program is passed to the following stages for further processing.

We found that \javac works quite well in localizing the error (at least the buggy line and finding the line is sufficient for our approach) \emph{if the program is free of nesting errors}. This is why we  apply \blkfix,  \emph{before} running \javac to localize and diagnose the error.

%
%


\subsection{Fixing Line Error: \linefix}
\label{sec:linefix}
\linefix uses a RoBERTa based pre-training + fine-tuning approach. 
RoBERTa derives from BERT, which
uses unlabeled text data to pre-train deep bidirectional representations of text by jointly conditioning on both left and right context in all layers of a deep transformer model~\cite{devlin2018bert} to perform simple, self-supervised tasks like filling in masked tokens.  
This model and training method effectively captures the statistics of token co-occurrences in very large corpora within the layers
of the transformer model. This pre-trained model learns excellent vector representations of code patterns in the higher layers
of the transformer; these learned vector
representations can be ``fine-tuned''
with just one additional output layer for specific tasks, and achieves state-of-the-art performance. For pre-training,
BERT uses two tasks: fill in masked out tokens using the context (also known as Masked language modeling, or ``MLM'') 
and predict the next sentence given the previous one (the ``NSP'' task). Liu \emph{et al.}'s 
RoBERTa (Robustly Optimized BERT Pretraining Approach) dominates BERT's performance~\cite{liu2019roberta}. Liu \emph{et al.} drop the NSP objective but  dynamically change the masking pattern used in the MLM of  BERT models. 

Pre-training + fine-tuning also works very well indeed for code. One can gather millions of unlabeled code tokens from open-source projects, conduct pre-training, and then fine-tune the model with a limited amount of labeled data to achieve state-of-the-art performance in different software engineering applications~\cite{feng2020codebert,kanade2020learning,biswas2020achieving,zhang2020sentiment} (albeit not yet for code syntax  repair). 
Since we are working on novice code correction and our objective does not involve any relation between two programs, such as Question Answering (QA) and Natural Language Inference (NLI), training on NSP is not beneficial. Furthermore, using a dynamic masking pattern to the training data helps the model achieve better performance in downstream tasks. Therefore, We use RoBERTa for pre-training and fine-tuning of the model. 

\vspace{0.1in}
\noindent{\bf Why pre-training?} As explained in the papers on BERT~\cite{devlin2018bert} and RoBERTA~\cite{liu2019roberta}, 
for natural language, and the very recent, but rapidly growing body of literature using pre-training for code~\cite{feng2020codebert,kanade2020learning,biswas2020achieving,zhang2020sentiment,  lu2021codexglue,roziere2021dobf, mastropaolo2021studying,jesse2021learning,9520296,guo2020graphcodebert,qi2021prophetnet,ahmad-etal-2021-unified}, pre-training is a way to exploit enormous volumes of data in a self-supervised fashion
to learn the statistics of token sequences, and capture patterns in a position-dependent vector notation. For our purposes, 
these pre-trained models are automatically ingesting patterns of syntax and identifier usage from vast quantities of
source code (around a billion tokens) and bringing all this knowledge implicitly to bear to the task of fixing errors
in syntax and identifier usage.

\begin{figure}[h!]
    \centering
    \includegraphics[scale=0.28, trim={0 1cm 0cm 0cm}]{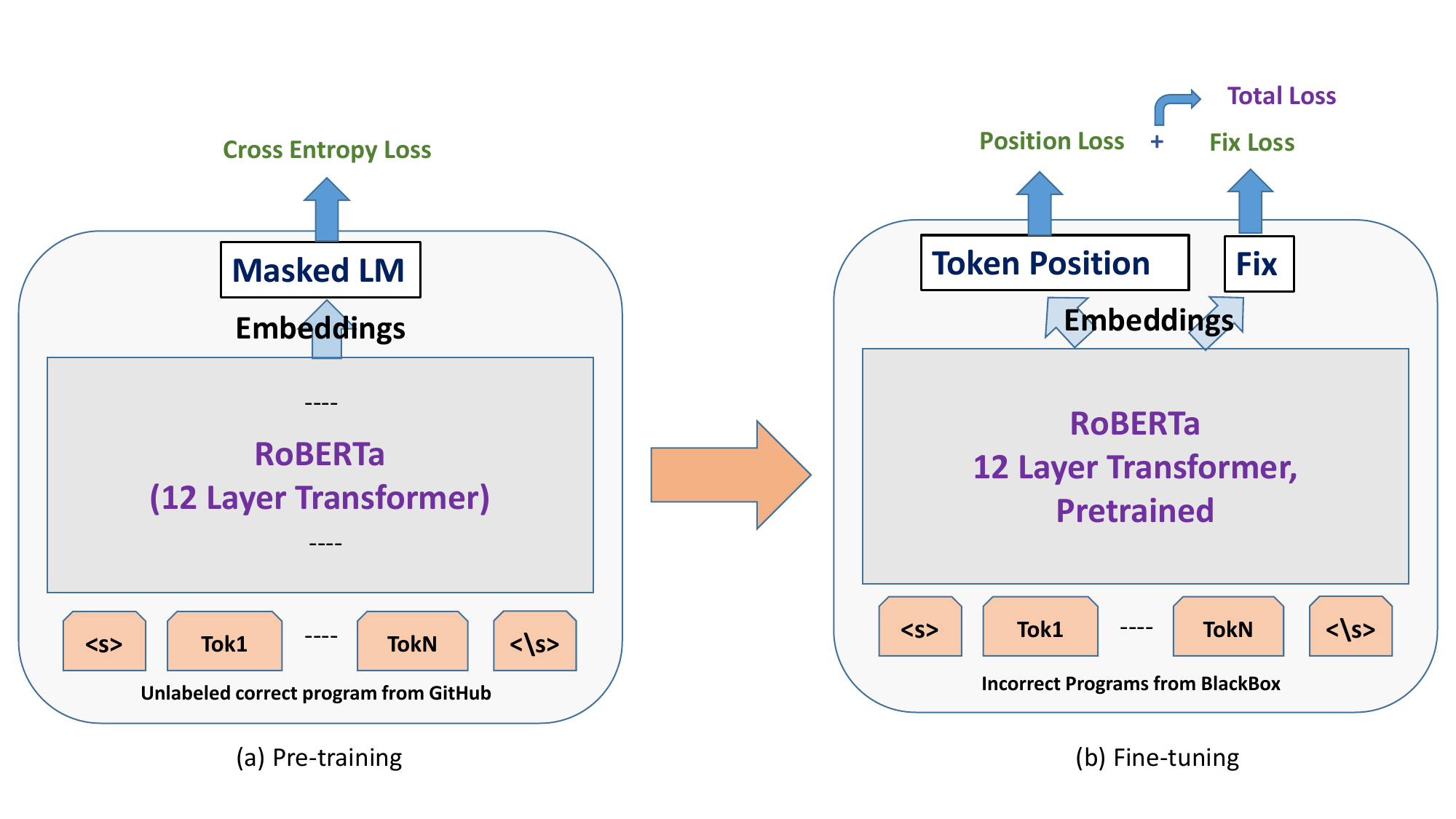}
    \caption{Pre-training and fine-tuning using RoBERTa.}
    \label{fig:roberta}
\end{figure}

\smallskip             
\noindent{\underline{\emph{Pre-training}}}
To generate the dataset for pre-training, we collected 5000 most starred Java projects from GitHub (since our end-goal
is to correct Java syntax errors). We tokenized the files, yielding 1.2 billion tokens for the pre-training. 
For the  MLM pre-training over code, we randomly  select 15\% of tokens, and replace with a unique token ${\mathtt{mask}}$. The loss here is the cross-entropy of the original masked token. Of the 15\% selected tokens, 80\% are replaced with a specific
marker ${\mathtt{mask}}$, 10\% are left unchanged, and a randomly selected token replaces the remaining 10\%. This training
method follows the standard RoBERTa protocol. 

\todone{I have gotten to this point of the section labelled linefix, and it has told me nothing about what you want to do with this pretrained model. I think from a reviewer persepective if you actually connect the task at hand with the need for a pre-trained model it would be helpful. I understand that there need to be certain details about roberta etc, but I would really tone it down and bring it all back to a SE context.}
The architecture is as shown in Fig.~\ref{fig:roberta}. The main RoBERTa model is in the central grey box, labeled ``RoBERTa'' in 
\ref{fig:roberta} (a) and \ref{fig:roberta} (b). The left side is the architecture when RoBERTA is being pre-trained; the last layer on top is the MLM, 
implemented as a softmax layer taking the RoBERTa embeddings as input, and produces an output token,   
The entire model is trained using cross-entropy loss. Our RoBERTa architecture 
consists of 12 attention layers, 768 hidden dimensions, and 12 self-attention heads in each layer. We applied Byte Level BPE (Byte Pair Encoding) tokenizer~\cite{karampatsis2020big} limiting the sub-token vocabulary size to 25K. \todone{say something about final MLM layer. Softmax output?}

We trained the MLM model using cross-entropy loss on two NVIDIA Titan RTX GPUs for five epochs with a batch size of 44 sequences and learning rate $5e-5$. When pre-training completed, our MLM model achieved a final loss corresponding to a perplexity of 1.46, (cross-entropy 0.546 bits)
 which is rather low; RoBERTa for natural language yields final losses around 3.68-4.0 perplexity (1.88 to 2 bits). 

%

\smallskip             
\noindent{\underline{\emph{Fine-tuning}}}
The fine-tuning step here is to train \linefix, a model that accepts an incorrect input line from a novice program,
(the line flagged by \javac as containing a syntax error) together with the text of the error itself, 
 and then generates a set of locations and edit commands, using multi-label classification layers, as explained below. 

For fine-tuning and then for evaluation, we used realistic novice programs with syntax errors and human-produced fixed versions. We used the exact dataset used by Santos \etal\cite{santos2018syntax}  and Ahmed \etal\cite{ahmed2021learning} from the Blackbox~\cite{brown2014blackbox} repository. This dataset contains 1.7M pairs, of erroneous and fixed programs. Both Santos \etal and Ahmed \etal primarily report their performance on programs with a single token error because a single edit can fix a large fraction of the programs (around 57\%). 
Therefore, for a fair comparison, we also initially focused our evaluation on single token errors and 
broke down our performance by token-length, as done 
by Ahmed \etal We selected a test set of 100K samples, with samples stratified by length, \todone{it might be late but I do not understand this "ten token" range}
from the full dataset for the evaluation. We divided the test dataset into ten token-length ranges (lengths of 1-100, 101-200, ... , and 900-1000 tokens), with each range having around 10K examples.
We prepare our fine-tuning dataset from the remaining examples. 


Since \blkfix handles long-range block-nesting errors, 
the \linefix stage is focused on those
errors unrelated to nesting. 
We discarded the programs with imbalanced curly braces from the training set, and after tokenization, we found around 540K examples to train the model. We used \javac (discussed in Section~\ref{javacnew}) to localize the error. The input to the model then is  the buggy line  indicated by \javac, appended with a special separator token (denoted {\tt <SEP>}) followed by the error message from \javac. Altogether, the maximum
input is 150 sub-tokens, which captures virtually all the input lines flagged as erroneous in our dataset. From this, the  pre-trained
RoBERTa model calculates positional embeddings for each subtoken; however, as with many RoBERTa-based classification tasks,
we use just the embedding of the first token. 

The desired output is the matching edits required to create the fixed version, as explained next. 

To make a complete fix, 
the model should produce one or more locations, and one or more  ``fix'', \emph{viz} edit commands. 
The fix has two parts: 
 i) the \emph{type} of fix (insertion, deletion, or substitute?) ii) the \emph{content} of  the fix (is it a specific keyword, delimiter, or any other token?). When the \emph{type} is a deletion, there is no \emph{content} required:  
if the model identifies the buggy token at position $x$ and recommends deletion, we just drop that token. For substitute operation, if the location is $x$ and the edit command is $substitute \rightarrow y$, we will replace the token at position $x$ with the token $y$. 
For insertion, if the command for position $x$  is $insert \rightarrow y$, we will add the suggested $y$ token at the $x+1$ position. 
For insertion at the start of the line, we use a special token. 
For example, consider the following buggy line from Fig.~\ref{fig:javac} (a).

\begin{smcodetabbing}
{\tt \textbf{\textit{public static void (String args[])}}}\\
\end{smcodetabbing}
\vspace{-0.4cm}

To fix this missing ``main'', \linefix should output the location ``3'' and the  fix ``$insert \rightarrow unk$'' (``main'' is an identifier). This ``unk'' will be coverted to ``main'' with another model. We will discuss it in Section~\ref{sec:UnkFix}. 

Our model's final layer consists of two distinct multi-label classification output layers, one which outputs one or more locations, another which outputs one or more fixes. 
The input to both these output layers, as explained above, is the RoBERTa embedding of the first token of the input. From this input, the
two separate multi-label classification output layers calculate the position(s), and fix(es).  Since most
(99\%) of the erroneous lines are 100 tokens are less, we output one or more positions (1-100) from the first output layer, and,
from the second output layer we generate 
one or more of 154 distinct possible fixes. 
We remind the reader that a multi-label classification task involves generating an output vector of class probabilities, where the classes
are non-exclusive. A single input might generate one or more class labels. 
In our case, we take all class labels in the output vector scoring above 0.5 as an assigned label. If none
of the classes are assigned a probability above 0.5, we just take the highest probability class label. 
In almost all cases, we have only one fix per line, so one position and one edit command are expected; however,
in rare cases, more than one position \emph{and} more than one edit command could be generated. In the former case,
we just apply the edit command at that position; in the latter case, which occurs very rarely, we try all combinations and return the first
edit combination that compiles. A somewhat more common case (for example with multiple missing delimiters, like ")"), we get one edit command
like $insert \rightarrow )$ 
and multiple locations, in which case, we just apply the same edit at all locations.


There are reasons for our choice of multi-label classification, rather than simply synthesizing the fixed output. 
Prior approaches~\cite{ahmed2021learning,gupta2017deepfix,santos2018syntax} used autoregressive\footnote{Autoregressive generation
conditions the generation of each token on previously generated tokens, and is used in machine-translation approaches.} 
code generation to synthesize repairs. Given the sizeable vocabularies in code, many complex dependencies  must
be accounted for when generating code tokens conditional on previous tokens, the original input tokens, and the
compiler error. 
We simplify the problem into a  multi-label classification task here; all that is required is to identify the token positions(s)
of the error, and the applicable edit commands. In the vast majority of cases, there is usually only a single change required
per line).  
This allows the model to learn, and rapidly reduce training loss and perform well under test. 
In addition, the multi-labeling approach (rather than auto-regressive
generation also allows us to handle repairs that require multiple fixes on the same line (example below, Fig~\ref{fig:multifix}). 
It's important to note that a single line can contain several token locations with errors, and distinct edit commands at each
position. 
Limiting the size of the set of possible fixes to 154 will limit the ability to fix identifier names; this is handled by including 
fix commands that insert and substitute to $unk$ in the output vocabulary of LineFix; these fixes are handled by 
a component is called 
\unkfix, which is described in \S~\ref{sec:UnkFix}. 
Note that dealing with multiple fixes on different lines is easily manageable.  If there are multiple positions, all with the same fix (like Fig.~\ref{fig:multifix}), one can just perform that fix at all the positions. However, for multiple positions and multiple fixes one needs to try all combinations until the \javac accepts with no errors.  We did not incorporate that to our code,  because:
\begin{enumerate}
\item Trying all possible combinations will slow down the entire process.
\item Two different errors in a line (even in a file) is very rare.  In the Blackbox data repository,  for example,  the  majority of files contain just a single syntactical error.  
 
 \end{enumerate}


\todone{here is the full description--- Since we are doing multi-label classification to allow the model to resolve more than one error in the same line, we use Binary cross-entropy with logits (BCEWithLogitsLoss) loss function instead of vanilla cross-entropy loss (CrossEntropyLoss). Binary cross-entropy loss assigns independent probabilities to the labels. Setting a threshold on the probability, we can easily find the output suggested by the model. We set 0.5 as a threshold. However, in a few cases model assigns less than 0.5 to all labels. In those cases, we take the label with maximum probability as output because the compiler already indicates the existence of the error. Our dataset mostly consists of the files having one error in the line. Therefore, almost in all cases, we get only one label from each output layer. If we get more than one from both output layers, we can try all possible combinations by replacing the buggy line with the javac-based compiler's help. }

The standard way to train multi-label classification layers is with binary cross-entropy loss (with logits), which is what we use for our fine-tuning. 
Since both the output layers are closely related to each other, we fine-tuned them simultaneously for 5 epochs. We collected the loss from each layer and added them to define the batch's final loss, and updated the model accordingly. Note that the same pre-trained model parameters
(from Fig.~\ref{fig:roberta} (a)) are used to initialize these; during fine-tuning, all parameters  in all layers are modified (Fig.~\ref{fig:roberta}). We use the Huggingface open-source
implementation of RoBERTa~\cite{huggingface} for both pre-training and fine-tuning. 

\begin{figure}
\begin{smcodetabbing}
{\tt \textit{\textbf{\textcolor{red}{-System.out.println(*);}}}}\\
{\tt \textit{\textbf{\textcolor{blue}{+System.out.println("*");}}}}\\
\end{smcodetabbing}
\caption{Example requiring two edits to fix} 
\label{fig:multifix}
\vspace{-.2cm}
\end{figure}

\noindent{\underline{\emph{Utilizing Compiler Diagnostics during Fine-tuning}}}
Apart from localizing the erroneous line, the compiler warning can boost the performance of the fine-tuning model. As an input sequence to the model, we tried two versions, \emph{i.e.,} with the warning, without warning. We observed a small but significant  improvement in line-level code fixing (detailed in Section~\ref{lnfixnew}). Consider the following code snippet from the Blackbox dataset. 
The variable ``bmr'' is declared twice, and the second declaration is invalid. Though the \javac  localizes the error correctly, it is really hard for the model to resolve this without any hint. Our model fails to fix this one when trained without the compiler message. However, with the compiler error message, our RoBERTa-based fine-tuned model can solve errors like this one by deleting the token ``double''.  This particular example is fixable with a modern IDE; however, it serves as a good illustration of how our model can use error messages. We remind the reader that in general we can handle numerous
examples that IDEs cannot. Several typical examples are included in the supplemental file \textcolor{blue}{\href{https://bit.ly/3CMM0TP}{https://bit.ly/3CMM0TP}}. 



\begin{smcodetabbing}
{\small\tt }\\
{\small\tt double bmr;}\\
{\small\tt /* some additional irrelevant lines */}\\
{\small\tt boolean isMale = male == 'M';}\\
{\small\tt if(isMale)}\\
{\small\tt double bmr = ((9.5 * wgt) + (5.0 * hgt) }\\
{\small\tt    \hspace{3cm}        + (6.7 * age) + 66.47);}\\

\end{smcodetabbing}

\todone{the error from javac is unclear from this listing}

\noindent \fbox{\begin{minipage}{26.3em}

\noindent{\scriptsize\bf Without Warning:} 

\noindent{\scriptsize\tt double bmr = ((9.5 * wgt) + (5.0 * hgt) + (6.7 * age) + 66.47);}

\noindent{\scriptsize\bf With Warning:}

\noindent {\scriptsize\tt double bmr = ((9.5 * wgt) + (5.0 * hgt) + (6.7 * age) + 66.47); <SEP> variable declaration not allowed here}

\end{minipage}}
\smallskip

\linefix works best with small sequences. Java is inherently verbose, and so sequence lengths are often beyond the model's capacity. Compiler diagnostics help us in two ways. Primarily, it helps us localize the error, and secondly, the message (even if imprecise) helps deep learning models fix the error. This claim is supported by a study (Yasunaga \etal~\cite{yasunaga2020graph}).

\subsection{Recovering Unknown Tokens: \unkfix}
\label{sec:UnkFix}
\todone{while I get it...you might want to highlight the place where these unks are put in and why}
Recall that LineFix output is restricted to 154 distinct fixes in the fine-tuning model. To deal
with edits (inserts or substitutes of identifiers, constants \emph{etc.}) outside of the limited vocabulary of edits, have an ``escape'' mechanism. 
Out of these 154, we included
two unique outputs $insert \rightarrow unk$ and $substitute \rightarrow unk$ to cover other changes.
To precisely identify these ``unk'' tokens, we use \unkfix, which reuses the masked-language model (MLM) we obtained during
pre-training. 
This masked language model can recover the $unk$ tokens if sufficient context is given. After getting the position information, we can collect sufficient tokens from the previous and following lines to fill the input buffer, 
and ask the pre-trained model to unmask the $unk$. Applying this approach, we could fix several $unk$-related program errors like the following ones where the LineFix predicts $insert \rightarrow unk$ and  $substitute \rightarrow unk$ for ``Item'' and ``Integer'', and then the MLM is able to locate them correctly. 

\begin{smcodetabbing}
{\tt }\\
{\tt \textit{\textbf{\textcolor{red}{-public void takeItem (item ) \{}}}}\\
{\tt \textit{\textbf{\textcolor{blue}{+public void takeItem (Item item ) \{}}}}\\
\end{smcodetabbing}

\begin{smcodetabbing}
{\tt \textit{\textbf{\textcolor{red}{-float number = float.parseInt(text);}}}}\\
{\tt \textit{\textbf{\textcolor{blue}{+float number = Integer.parseInt(text);}}}}\\
\end{smcodetabbing}

\noindent Note that though we designed \unkfix primarily for identifiers, it can potentially handle other tokens, including values. 
            
\todone{I assume this is the end of the list of errors that can be fixed. UnkFix fixes issing tokens (classes/method names) and line fix probably fixes a certain class of issues as well. It would be best to provide an overview of all classes of issues that synfix can fix.}             

\subsection{Integrating SYNSHINE into VSCode}
\label{sec:synfix}

%
%
%
%

To make \newname more broadly accessible, we have made it available within a popular IDE. 
We have initially chosen VSCode since it's widely available, free for students\footnote{\url{https://visualstudio.microsoft.com/students/}}, and well-documented; in the  future, we will  incorporate \newname into other IDEs. The source code for the integration is available in 
our replication package. 
\todone{I like the video, but one additional thing could be to hover over the error to show the viewer what current IDE feedback is on each error that synfix fixes.}
A demo video is viewable: \textcolor{blue}{\href{https://youtu.be/AR1nd2PJczU}{https://youtu.be/AR1nd2PJczU}}.

In this VSCode integration, we desired fast response times, and wanted to avoid the requirement for a GPU, since many
novices may not have a GPU. So for the \newname deep learning model, we just used  CPU floating point operations; 
to avoid having to reload the (very large)
model for each repair request, we wrapped the \newname model within a ``correction'' server, which services HTTP requests from the IDE. 

The IDE triggers
a request to \newname when the user requests a fix suggestion. 
When \newname is triggered, VSCode looks for the active text editor and extracts the (erroneous) code content from there. After getting the content, VSCode sends an HTTP request to the code correction server. Models are pre-loaded in the correction server, so that it can immediately service requests. In this server, the code goes through our proposed pipeline presented in Fig.~\ref{fig:pipeline}, and the code returns to the editor after finishing all the steps. Now we have two versions of the code, i.e., the buggy code and the corrected version. We highlight the difference and present both versions 
to the user and allow them to accept or reject the solution.

Note that the demo presented on the link mentioned above was captured on a machine without any GPU. We observe that \newname can operate on a CPU and is quite fast at generating the solution even though the models were trained on GPUs. Just to get a sense of the delay, we randomly chose 200 erroneous programs of various lengths from  our dataset, and measured the response time (time from the ``\newname'' button press to the time the fixed
code is received back). The average response time is 0.88 seconds (standard deviation 0.49s, maximum 2.2s). While this by no means instantaneous, we can still provide a fix for a syntax error virtually always within a second or two, potentially saving the novice and instructor's time. Our approach to 
integrating
 \newname into VSCode 
 thus arguably attenuates the need for expensive GPUs,   and facilitates the use of the deep learning model in CPU-only machines.  The CPU we used for the experiment is ``AMD Ryzen 7 2700X''. The code correction server occupies 1.765 GB of the memory. 
 
SynShine's response time is significantly lower than the time needed by a programmer to fix the program. Brown and Altadmri divided the mistakes that occurred in the Blackbox repository into 18 different classes, where 11 of them are syntactical errors~\cite{brown2017novice}. The programmers take 13-1000 seconds (median) to fix the mistakes~\cite{brown2017novice}.   
Our model, on the other hand,  takes less than a second on average to process the files and suggest a fix. 
 
    

\section{Evaluation \& Results}
\label{result}
\todone{If we use different sample sizes in the different evals, we should be clear about it,
and also justify}
In our evaluation, we compare our work with several baselines: Santos \emph{et al.}, DeepFix , \oldname, and
SequenceR. The original DeepFix~\cite{gupta2017deepfix} used a GRU based
RNN encoder-decoder translation model, which takes an entire program (with syntax
error) as input, and produces a fix.  For baselining their \oldname tool, 
\todone{I would just saw "we" used this just like Ahmed et al.} Ahmed \emph{et al.} 
used two versions of DeepFix, one (``short'') trained on error-fix pairs upto 400 tokens long and
another (``long'') trained on error-fix pairs upto 800 tokens long.  Another approach, SequenceR~\cite{chen2019sequencer}
has reported success in fixing
\emph{semantic} errors, when provided with fault localization; 
it is also adaptable for syntax errors. SequenceR differs from DeepFix in a few ways: it uses
a separate fault localizer, and also incorporates a copy mechanism. 
We describe the intricacies in full detail later. 
Ahmed \emph{et al.}'s  \oldname program 
used a 2-stage transformer-based lenient parser, as described above. 
Our approach combines several techniques: pre-training, compiler-based reporting, and fine-tuning
with novice data. 

Below, we present summary top-1 accuracy results, evaluated over a random sample of 100,000 examples of length upto 1000 tokens, with single-token
errors, taken from the Blackbox dataset. The detailed result is presented in Table~\ref{compare1}. We follow the lead of the first paper in the area~\cite{santos2018syntax} in
this table, reporting performance for single-token errors, which constitute 57\% of the data in Blackbox. We report
the numbers for more complex errors below. 
\todone{might want to highlight your own performance in the table}
\begin{table}[h]

\label{pretrain}
\centering
\resizebox{0.88\columnwidth}{!}{%
\begin{tabular}{|c|c|c|c|c|c|}

\hline
\multirow{2}{*}{\textbf{\begin{tabular}[c]{@{}c@{}}Santos \\ \etal~\cite{santos2018syntax}\end{tabular}}} & \multirow{2}{*}{\textbf{\begin{tabular}[c]{@{}c@{}}DeepFix\\ (short)\end{tabular}}} & \multirow{2}{*}{\textbf{\begin{tabular}[c]{@{}c@{}}DeepFix\\ (long)\end{tabular}}} & \multirow{2}{*}{\textbf{SequenceR}} & \multirow{2}{*}{\textbf{\oldname}} & \multirow{2}{*}{\textbf{\newname}} \\
                                                                                   &                                                                                     &                                                                                    &                                     &                                            &                               \\ \hline
\multicolumn{1}{|l|}{46.00\%}                                                             & \multicolumn{1}{l|}{63.25\%}                                                        & \multicolumn{1}{l|}{62.14\%}                                                       & \multicolumn{1}{l|}{56.89\%}        & \multicolumn{1}{l|}{56.91\%}               & \multicolumn{1}{l|}{\bemph{74.89\%}}  \\ \hline
\end{tabular}
}
\caption{{\small\em Summary Results: Santos \etal performance is as reported by them; we measured the others}}
\vspace{+0.0cm}
\end{table}
As can be seen, \newname achieves a substantial performance boost, over all the prior approaches, elevating the performance further 
and providing us with the motivation to build it into a popular IDE to make
it more widely available.  Here below, we evaluate the performance in more detail, comparing \newname with
the closer competitors (we exclude Santos \etal from this comparison) 
and also examine the contributions
of our various stages to the significant overall improvement. We begin with an evaluation of the
effect of program length on performance, then we consider the effect of the various components of \newname. 
Finally, we breakdown the performance of \newname in repairing various categories of syntax errors.

\subsection{Fixing shorter \& longer programs}

\todone{I may have missed it, but I think this is the first time BlueJ is introduced.}
\todone{can you explain what longer problems you can solve that no one else can? or better yet what exactly is a long dependency problem? misspelling of a field that is 800 tokens before its actual use?}
\begin{table*}[h]

\centering
\resizebox{0.90\textwidth}{!}{%

\begin{tabular}{llllllllll}
\hline
\multicolumn{1}{c}{\multirow{2}{*}{\begin{tabular}[c]{@{}c@{}}Token \\ Range\end{tabular}}} & \multicolumn{1}{c}{\multirow{2}{*}{\begin{tabular}[c]{@{}c@{}}Percent of\\ Overall Data\end{tabular}}} & \multicolumn{1}{c}{\multirow{2}{*}{\begin{tabular}[c]{@{}c@{}}DeepFix\\ (short)\end{tabular}}} & \multicolumn{1}{c}{\multirow{2}{*}{\begin{tabular}[c]{@{}c@{}}DeepFix\\ (long)\end{tabular}}} & \multicolumn{1}{c}{\multirow{2}{*}{SequenceR}} & \multicolumn{1}{c}{\multirow{2}{*}{\oldname}} & \multicolumn{4}{c}{\newname}                                                                                                                                                                                                                                           \\ 
\multicolumn{1}{c}{}                                                                        & \multicolumn{1}{c}{}                                                                                   & \multicolumn{1}{c}{}                                                                           & \multicolumn{1}{c}{}                                                                          & \multicolumn{1}{c}{}                           & \multicolumn{1}{c}{}                                  & \multicolumn{1}{c}{\begin{tabular}[c]{@{}c@{}}By \\ \blkfix\end{tabular}} & \multicolumn{1}{c}{\begin{tabular}[c]{@{}c@{}}By \\ \linefix\end{tabular}} & \multicolumn{1}{c}{\begin{tabular}[c]{@{}c@{}}By\\ \unkfix\end{tabular}} & \multicolumn{1}{c}{Total}   \\ \hline
            
1-100                                                                                         & 31.01\%                                                                                                 & 76.71\%                                         & 73.72\%                                        & 59.21\%                                         & 65.16\%                                                & 21.01\%                                                                     & 58.86\%                                                                    & 2.41\%                                                                   & \bemph{82.28\%}                      \\
101-200                                                                                       & 29.43\%                                                                                                 & 69.98\%                                         & 67.15\%                                        & 57.21\%                                         & 60.24\%                                                & 17.53\%                                                                     & 58.98\%                                                                    & 1.96\%                                                                   & \bemph{78.47\%}                     \\
201-300                                                                                       & 15.25\%                                                                                                 & 63.27\%                                         & 60.29\%                                        & 55.40\%                                         & 54.47\%                                                & 14.35\%                                                                     & 56.00\%                                                                    & 1.93\%                                                                   & \bemph{72.28\%}                       \\
301-400                                                                                       & 8.56\%                                                                                                  & 53.71\%                                         & 54.02\%                                        & 54.64\%                                         & 50.01\%                                                & 10.18\%                                                                     & 54.45\%                                                                    & 1.89\%                                                                   & \bemph{66.52\%}                       \\
401-500                                                                                       & 5.51\%                                                                                                  & 42.17\%                                         & 45.47\%                                        & 54.54\%                                         & 46.19\%                                                & 7.71\%                                                                      & 54.00\%                                                                    & 1.88\%                                                                   & \bemph{63.59\%}                       \\
501-600                                                                                       & 3.63\% & 32.84\%                                         & 39.78\%                                        & 54.47\%                                         & 42.81\%                                                & 5.95\%                                                                      & 53.83\%                                                                    & 2.19\%                                                                   & \bemph{61.97\%}                       \\
601-700                                                                                       & 2.17\%                                                                                                  & 23.76\%                                         & 33.02\%                                        & 54.35\%                                         & 38.07\%                                                & 3.80\%                                                                      & 53.62\%                                                                    & 2.10\%                                                                   & \bemph{59.52\%}                       \\
701-800                                                                                       & 1.90\%                                                                                                  & 17.10\%                                         & 26.57\%                                        & 53.78\%                                         & 35.35\%                                                & 3.04\%                                                                      & 51.98\%                                                                    & 2.65\%                                                                   & \bemph{57.67\%}                       \\
801-900                                                                                       & 1.34\%                                                                                                  & 11.43\%                                         & 22.88\%                                        & 55.56\%                                         & 32.24\%                                                & 2.20\%                                                                      & 52.84\%                                                                    & 2.19\%                                                                   & \bemph{57.23\%}                       \\
901-1000                                                                                      & 1.19\%                                                                                                  & 8.80\%                                          & 17.94\%                                        & 53.87\%                                         & 29.62\%                                                & 1.27\%                                                                      & 51.63\%                                                                    & 2.10\%                                                                   & \bemph{55.00\%}                       \\ \hline
\multicolumn{2}{l}{Overall}                                                                                                                                                                           & \multicolumn{1}{l}{63.25\%}                    & \multicolumn{1}{l}{62.14\%}                   & \multicolumn{1}{l}{56.89\%}                    & \multicolumn{1}{l}{56.91\%}                           & \multicolumn{1}{l}{15.56\%}                                                & \multicolumn{1}{l}{57.22\%}                                               & \multicolumn{1}{l}{2.11\%}                                              & \multicolumn{1}{l}{\bemph{74.89\%} } \\ \hline
\end{tabular}

}
\caption{{\small\em Baselining \newname against prior work on syntax error correction. SequenceR was provided with \javac localization.}}
\label{compare1}
\end{table*}

Table~\ref{compare1} baselines  the relative performance of \newname against prior work, broken down by length, in categories. The rows are different
length ranges of  programs. The second column is the fraction of the Blackbox programs falling in this length range. 
The next several columns are are baselines from prior work: first two are DeepFix (short) trained  on shorter error-fix pairs (upto 400 tokens long), DeepFix (long) trained on pairs up to 800 tokens long. The next two are SequenceR, trained on all pairs in the training set, 
and \oldname, trained exactly provided in Ahmed \emph{et al.}'s scripts. Finally, on the last column we have our results from \newname;
the 3 columns to the right of the \newname column represent the contributions of our 3 components. 
As can be seen our overall performance exceeds the performance of all the others in every length category, and on the entire
sample significantly improves on all of them. Before we examine the numbers in detail, we first present some relevant details 
on how we measured them. 

All evaluations were done on a very large, randomly chosen, representative sample of 100,000 error-fix pairs from Blackbox that were not seen during
training by any of the models. The percentages shown in the second column, and the overall performance numbers (all numbers are \emph{top-1 accuracy}) 
are thus robust estimates of actual performance on programs up to 1000 tokens long, which constitute around 95\%  of the Blackbox data. An additional evaluation
on a random sample of the entire dataset is reported below. \todone{need to explain this two steps further. why didn't we include 1000+token programs in this set?} \todone{1. Trying deepfix over 1000+ seems expecting too much from DeepFix. 2.EMSE paper didn't do it. We are following that paper's strategy}
DeepFix (short), DeepFix (long), and \oldname were all trained and evaluated using the scripts made available in the replication package of Ahmed \emph{et al.}~\cite{ahmed2021learning} and Gupta \emph{et al.}~\cite{gupta2017deepfix}. 

SequenceR~\cite{chen2019sequencer} had to be retrained
for syntax error correction: Chen \emph{et al.} originally developed SequenceR for fixing \emph{semantic} bugs, viz., test failures. 
It uses the OpenNMT translation framework~\cite{klein2017opennmt} and thus had to 
be trained using bug-fix pairs. SequenceR assumes that the precise location of the bug was
known via fault-localization; the training pairs consisted of a) the buggy region of code, bracketed within {\small\tt <start\_bug> $\ldots$ <end\_bug>} markers, 
augmented with sufficient context (preceding and succeeding tokens)
to make up 1000 tokens of input b) and the corresponding fix, which is the region including the changed code, upto a maximum of 100 tokens; longer fix regions will fail
(this almost never happens in our setting). They used an RNN sequence-to-sequence encoder-decoder model that uses LSTM for the recurrent nodes,
and incorporates a copy mechanism to enable the model to generate specific local variables, \emph{etc.} in fixes. We used the code provided by
Chen \emph{et al}, and trained the model using Blackbox data; we used the \javac compiler to find the error location, and created training/test
pairs using the \javac indicated location (with context), together with the corresponding novice fix. In our case, since most novices' programs are shorter
than 1000 tokens, we provided the entire novice program as context. Once SequenceR is trained, it can generate fixes, given
the  novice program with error, with location indicated as above. However, SequenceR cannot insert or delete entire lines, so it cannot
 fix many nesting errors (for example, by inserting or deleting a line with a single "\{" or "\}" delimiter). 

Our overall accuracy ranges between 55\% to 82\%, and always outperforms  DeepFix long (18\%-74\%), and short (9\%-77\%), SequenceR  (54\%-59\%) and \oldname (29\%-65\%).  Both SequenceR and \newname benefit from the error location provided by \javac. 
By improving on prior work at every range, on the entire representative 100,000 sample, \newname achieves significant gains in \emph{overall performance} 
(bottom line) over the state of the art. 
Two factors contribute to this improvement: i) javac-based error localization and ii) robustness of \linefix and \unkfix. javac-based error localization enables a more
selective LineFix+UnkFix to the most likely errorful code, thus reducing false positives; Ahmed \emph{et al.}'s \oldname attempts corrections throughout the program,
resulting in more mistaken corrections.  The robustness of \linefix and \unkfix  is really boosted by the 
 pre-training + Fine-tuning strategy;  we explore  the relative benefits of this step further below.

Table~\ref{compare1}, in  columns under the \newname header, also shows relative contributions of the components of \newname.
First stage is {\sc \blkfix} borrowed from \oldname. 
About 20\%-25\% programs, regardless of length have nesting errors. \blkfix's accuracy decreases with program length, and we observe that the contribution of \blkfix is low after 700 tokens. However, for the other 75\% to 80\% programs without nesting errors,   \linefix \& \unkfix perform pretty consistently. 
Finally, we note that 1-1000 tokens cover about  95\% of the overall data.
To observe the performance of \newname on the overall distribution, including programs over 1000 tokens long, 
we test it on 5000 random samples. We found that our model can repair 75.36\% of the programs,
and as before, comfortably exceeds performance of prior tools. 
Note that if the \blkfix model has already fixed the curly braces and there is no other error, \javac will not produce any error message, and \linefix will not process that. 
Note that we always compare the end-to-end tokens of the reference and the model’s proposed sequence; if needless ``over'' fixes are applied,  that will be counted as wrong.
Moreover, none of the fixes are credited twice. 
If the model is fixed by \unkfix,  it alone receives credit;
we did not count it in the \linefix column.  Likewise,  we credited a sample in the \linefix column, if it is completely fixed by \linefix and does not receive any help from \unkfix.

We also applied our model on files that required 2 and 3 edits to fix the program and observed 29.4\% and 14.4\% accuracy, which is much higher than the reported accuracy by Ahmed et al. (19\% and 9\%).    Finally, we note that Ahmed \etal report on a blended strategy where shorter uncompilable programs
could be sent to  DeepFix and longer ones to \oldname, thus obtaining better performance than either at all lengths. 
A similar strategy could be employed here, blending
\newname with other models, trying all the proposed solutions,  and picking the ones that compile.  
However we didn't implement this approach: we just integrated \newname into VSCode since it performs quite well at all lengths
on its own, and avoids the need to load and run many models, and try repeated compiles.


\subsection{{\sc LineFix:} The role of Compiler Errors}
\label{lnfixnew} 
\newname differs from both versions of DeepFix, and SequenceR, because it's multistage; it differs from
 \oldname  mainly because of the two new components, {\sc LineFix}
and \unkfix. We simply reused the {\sc \blkfix} component made available by Ahmed \emph{et al.}~\footnote{https://zenodo.org/record/4420845},
and find performance very similar to that reported by them for this component. The improvements  reported in Table~\ref{compare1} clearly arise from our two new
components. We now focus in on \linefix and evaluate how it contributes to overall performance. \linefix's task is to take an input line flagged as 
a relevant syntax error (by \javac), together with the actual error, and then output a position, and an editing hint (insert, substitute, delete). \linefix improves upon the {\sc FragFix} stage of \oldname
in two ways: first, it uses pretraining+finetuning, and second, it also takes the syntax error message from \javac as an additional input. 
The value of pre-training has been extensively documented for code-related tasks~\cite{feng2020codebert,kanade2020learning,biswas2020achieving,zhang2020sentiment,  lu2021codexglue,roziere2021dobf, mastropaolo2021studying,jesse2021learning,9520296,guo2020graphcodebert,qi2021prophetnet,ahmad-etal-2021-unified}, so we focus here on the effect of providing compiler errors. 
Note again that \linefix has two tasks: \emph{Localize} the token to be replaced, and and output
an editing command with the correct \emph{Fix}. We evaluate the impact of compiler warnings using 10,000 randomly chosen erroneous lines, 
of various lengths, each taken with and without the compiler syntax error messages. Since we're evaluating fixing capability on single erroneous lines,
rather than entire programs, the numbers reported below are higher than in Table~\ref{compare1}.

\todone{so...this breakdown of linefix is a bit confusing. In the previous table I see that it averages 57.2\% accuracy and now it is 89\%? you might want to delve into the difference in the test setting.}
 
 \begin{table}[h]

\centering
\resizebox{\columnwidth}{!}{%
\renewcommand{\arraystretch}{1.4}

\begin{tabular}{llllllcc}
\hline
\multicolumn{1}{c}{\multirow{2}{*}{With compiler error? 
}} & \multicolumn{3}{c}{Localization} & \multicolumn{3}{c}{Fix}     & \multicolumn{1}{c}{Complete Correction (Location+Fix)} \\ 
\multicolumn{1}{c}{}                                 & \multicolumn{3}{c}{F-Score}      & \multicolumn{3}{c}{F-Score} & \multicolumn{1}{c}{Accuracy}                           \\ \hline
No                                                     & \multicolumn{3}{l}{90.75\%}       & \multicolumn{3}{l}{92.41\%}  & 86.71\%                                                 \\ 
\multicolumn{1}{l}{Yes}                              & \multicolumn{3}{l}{93.58\%}      & \multicolumn{3}{l}{93.18\%} & \multicolumn{1}{c}{89.39\%}                            \\ \hline
\end{tabular}

}
\caption{Impact of using compiler error}
\label{pretrain}
\end{table}

Table~\ref{pretrain} presents the impact of using the syntax error message in our tool. 

We gain around 2.7\% improvement in overall accuracy using the compiler
error message. 
We also see improvements
on both Localization and Fix f-scores by providing the compiler message along with the  the erroneous line (row 1 \& 2). The improvement is more for the Localization than for the Fix.  We tested the statistical significance of all differences, using Binomial difference of proportions test on
a trial sample of 10,000; we then corrected the p-values
using Benjamini-Hochberg. The improvements observed when using compiler error message
for overall accuracy and fix location f-score are \emph{highly} significant ($p < 1e-9$); however, the f-score for the fix \emph{per se} are only significantly
improved ($ 0.01 < p  < 0.05$). \todone{im gonna believe your stats...but the only 2.3\% of improvement suggests that the difference is not that pivotal so have you really double checked this?}This suggests that the compiler error message is of highly significant help in providing our model with information
required to locate the precise token that needs to be edited, and somewhat less so to identify the precise edit that is required. It is
\emph{very important} to note however, that the \javac compiler is of crucial help in locating the \underline{line} where the error
is located. This above study also shows that the actual error message \emph{per se} helps our model locate the \emph{token} within that line that needs to be edited. 

\smallskip             
We present an illustrative example of how compiler error messages help. 
Sometimes the compiler warnings are very precise, \emph{e.g.,} when semicolons or other punctuations are to be inserted. In such
cases, it may appear that the task is quite simple, and the model is simply ``translating'' the error into a fix. 
We sampled  50 programs and observed how many of them can be fixed just by reading the comments. 
We observed that in roughly 60\% cases, the \javac warning is not that helpful, and the model learns to respond
in fairly nuanced ways to address  the error. 
Consider the following repair that \linefix correctly achieves.

\begin{smcodetabbing}
{\tt \textit{\textbf{\textcolor{red}{-return s == reverse ( String s ) ;}}}}\\
{\tt \textit{\textbf{\textcolor{blue}{+return s == reverse ( s ) ;}}}}\\
\end{smcodetabbing}

\javac \emph{per se} not helpful: it produces an error message suggesting 
to insert ``{\small\tt )}'' after ``{\small\tt String}''.  \linefix learns to ignore such messages, 
and instead correctly omits the token ``String''. 
Therefore, the model is not just ``translating'' the message from \javac into a fix; 
The high capacity of the model, enriched by pre-training and fine-tuning, is deployed
to leverage the often incorrect, imprecise message from \javac into a good fix. 
Depending on the error, it can resolve a very imprecise message from \javac. 
Indeed, quite often the same error message from \javac can lead the model to 
provide very different (correct) fixes.

\subsection{When \newname fails, and when it works.} 
We now examine in further detail the cases where \newname works correctly, and where it does not. To be conservative, 
we have defined as a ``failure'' any fix  \emph{not exactly the same} as the one recorded in the Blackbox dataset; note that a) the fix recorded in Blackbox is created by an actual human
user, and also b) the recorded fixes always compile without error. We start with an examination of the cases where \newname  \emph{fails} to produce
a correct fix, as per our conservative definition, 
and then examine in detail the \emph{diversity} of fixes that it does provide.

\begin{table}[h]

\centering
\resizebox{.90\columnwidth}{!}{%

\begin{tabular}{lccc}
\hline
\multicolumn{1}{c}{Length} & \multicolumn{1}{c}{\begin{tabular}[c]{@{}c@{}}Overall \\ Compilability\\ of fixes \end{tabular}} & \multicolumn{1}{c}{\begin{tabular}[c]{@{}c@{}}Fixes \\ Exactly Matching\\ Blackbox \end{tabular}} & \multicolumn{1}{c}{\begin{tabular}[c]{@{}c@{}}Compilability \\ for \\ non-matching cases \end{tabular}} \\ \hline
1-100                             & 90.18\%                     & 82.28\%                                                                    & 44.58\%                                                                                               \\
101-200                           & 86.13\%                    &78.47\%                                                                    & 35.58\%                                                                                               \\
201-300                           & 79.33\%                      &72.28\%                                                                       & 25.43\%                                                                                               \\
301-400                           & 73.35\%                   &66.52\%                                                                          & 20.40\%                                                                                               \\
401-500                           & 70.14\%                 & 63.59\%                                                                            & 17.99\%                                                                                               \\
501-600                           & 67.83\%               &61.97\%                                                                              & 15.41\%                                                                                               \\
601-700                           & 65.92\%              &59.52\%                                                                               & 15.81\%                                                                                               \\
701-800                           & 64.00\%            &57.67\%                                                                                 & 14.96\%                                                                                               \\
801-900                           & 63.32\%             &57.23\%                                                                                & 14.24\%                                                                                               \\ 
\multicolumn{1}{l}{901-1000}  &60.76\%  & 55.00\%                                                                    & 13.00\%                                                                          \\ \hline
\end{tabular}

}
\caption{Compilability of \newname}
\label{compile}
\end{table}

\begin{table}[th!]

\centering
\resizebox{.8\columnwidth}{!}{%

\begin{tabular}{lcc}
\hline

\multicolumn{1}{c}{Category} & \multicolumn{1}{c}{\begin{tabular}[c]{@{}c@{}}Prevalence of \\ Error Category\end{tabular}} & \multicolumn{1}{c}{\begin{tabular}[c]{@{}c@{}}Fix Accuracy\\  (in \%)\end{tabular}} \\ \hline
Keyword                        & 5.04\%                                                                                  & 70.64\%                                                                          \\
Operator                       & 5.87\%                                                                                  & 77.73\%                                                                          \\
Delimiter                      & 80.37\%                                                                                 & 81.60\%                                                                          \\ 
\multicolumn{1}{l}{Other}    & 8.72\%                                                            & 60.94\%                                                     \\ \hline

\end{tabular}

}
\caption{Performance of \newname over diverse error categories}
\label{robust}
\end{table}

\smallskip             
\noindent{\underline{\emph{Fix Failures}}}
Despite our over-conservative definition of ``failure'', 
sometimes \newname can generate a solution that differs from the user-intended solution but 
is still compilable with our javac-based compiler. 
In some cases, the solution is even semantically correct. 
As an illustration, in Table~\ref{ex}, examples 1, 2 \& 3 are fixes generated by \newname that 
not only compile without error, but are also semantically correct. 
By contrast, the last example in Table~\ref{ex} is not semantically correct but compilable.
Ideally, we'd like to characterize how often \newname finds fixes that are not only compilable,
but also semantically correct. 
The \emph{compilability} of a fix that differs from the user's fix  recorded in Blackbox
can be determined automatically, and at scale (by just compiling!) and we report it below;
however, the \emph{semantic correctness} of a fix that differs from a user's fix  requires manual
examination, and is not practical to do at a large scale. We try to characterize these to some extent
by examining a small sample. 

\todone{someone really is gonna complain that the examples are cherry picked...where is the real dirt? especially the first 3 ones where you are over penalizing yourselves. toss a curveball and show a catastrophic failure too...}

\begin{table*}[h]

\centering
\resizebox{\textwidth}{!}{%

\begin{tabular}{|l|l|l|l|}
\hline
\multicolumn{1}{|c|}{Seq No} & \multicolumn{1}{c|}{Buggy Line}                        & \multicolumn{1}{c|}{Model}                           & \multicolumn{1}{c|}{Original Fixed}                  \\ \hline
1                            & int i = ( ( int ) ( Math . random ( ) * 3 ) ; & int i = ( ( int ) ( Math . random ( ) * 3 ) ) ; & int i = ( int ) ( Math . random ( ) * 3 ) ; \\ \hline
2                            & int userInt\_1 , int userInt\_2 ;                      & int userInt\_1 ; int userInt\_2 ;                    & int userInt\_1 , userInt\_2 ;                        \\ \hline
3                            & System . out . print ( ( " Hello, world. " ) ;         & System . out . print ( " Hello, world. " ) ;         & System . out . print ( ( " Hello, world. " ) ) ;     \\ \hline
4                            & System . out . println ( " sum = " + ( sum + ) ) ;     & System . out . println ( " sum = " + ( sum ) ) ;     & System . out . println ( " sum = " + ( sum + 5 ) ) ; \\ \hline
\end{tabular}

}
\caption{Examples showing the compilability of the model}
\label{ex}
\end{table*}

Table~\ref{compile} presents the overall compilability of the solutions.  The second column is the overall \emph{compilability} of the generated fix. This is calculated as  the fraction
of the number of attempted fixes, that actually results in a successful compilation. The third column is the proportion of fixes that we deem correct, based
on \emph{exact} match with the fix recorded in Blackbox (the numbers will match shown in the rightmost column of Table~\ref{compare1}). As can be seen, we
record many compilable cases as incorrect. The last column in Table~\ref{compile} shows the proportion of apparent failures that are actually compilable: as an illustration, for programs up to 100 tokens long, 
about 45\% of the cases that we record as an incorrect fix, in fact compile correctly. Depending on length, between 13\% and 45\% of the fixes
we classify as failures are actually compilable. Table~\ref{ex}, examples 1,2,3,4 are exactly such fixes. 


Now what proportion of these ``compilable  failures'' are actually semantically correct? 
To get a (very) rough estimate of this, 
we did a small manual study. We randomly collect 50 cases where the model generates a compilable fix, that fails to match the user fix recorded in Blackbox. 
We found that about 18\% of programs are semantically correct. 

To summarize: even in our very conservative evaluation, \newname produces the same
fixes as recorded by a human in a sizable fraction (roughly 75\%)  of errors in our novice dataset;  
an examination of \newname's failures suggests that 
it could possibly be helpful in some additional cases.

\smallskip             
\noindent{\underline{\emph{Fix Diversity}}}
What kinds of errors does \newname fix? In our dataset, 
about 80\% of the errors are related to delimiters, and even solving only those would make a significant dent. 
However, the novices make syntax errors in using 
keywords, operators, identifiers, and numbers; sometimes they introduce illegal spaces, declarations, characters, \emph{etc}.  We examined
how \newname performs with respect to different types of errors.  For convenience, 
we divided the error into four major categories- keywords (all Java keywords), delimiters (e.g., semicolon, comma, parentheses, braces, brackets), operators (all Java operators), and others (identifiers, literals, and anything that falls outside the first three categories). To do categorization, we followed two rules.
Errors that required substitutes or inserts belonged to the category of the substituted or inserted token; errors that required
deletion belonged to the category of the deleted token. Thus if an error required a semicolon to be inserted, it was in the ``delimiter'' category; 
if an error required an extra ``if'' keyword to be deleted, it was in the ``keyword'' category. 

%
%
%

We randomly sampled a 5K  test dataset, and determined the error category prevalence in this dataset; see Table~\ref{robust},  first column, for
the prevalence of errors in various categories. Delimiter errors dominate,
and thus our model learns to fix those best (81.6\% accuracy); however, 
it performs well in other categories (60\%-78\% accuracy). The take-away  from this analysis is that \newname
performs reasonably well at a wide range of syntax errors.

\section{Related Work}
\label{rework}

The most closely related works are DeepFix~\cite{gupta2017deepfix}, \oldname~\cite{ahmed2021learning}, and Santos \etal~\cite{santos2018syntax} which we have discussed above. We also discussed SequenceR~\cite{chen2019sequencer}. We
have compared \newname to all of these. 

Gupta et al.~\cite{gupta2018deep} applied reinforcement learning to a very similar dataset like DeepFix~\cite{gupta2017deepfix}. It utilizes total count of compiler errors as a part of the reward mechanism. However, RLAssist~\cite{gupta2018deep} shows only a very minor improvement over DeepFix~\cite{gupta2017deepfix}, and also it takes the whole program as input. Therefore, we did not re-implement RLAssist~\cite{gupta2018deep}. Though RLAssist~\cite{gupta2018deep} looks into compiler errors but it does not directly uses the error messages as we do.  DeepDelta~\cite{mesbah2019deepdelta} is another approach that fixes compiler errors but mostly identifier name-related errors, not syntax errors. DeepDelta~\cite{mesbah2019deepdelta} was developed and tested on code from professional developers at Google. The authors also assume that precise knowledge of the location will be given to the program.
Yasunaga \etal~\cite{yasunaga2020graph,yasunaga2021break} introduce two compiler-dependent approaches to fix \cp program: DrRepair that utilizes \cp compiler warnings with a graph-based self-supervised approach, and BIFI that applies two models ``critic'' and ``fixer'' to fix the programs.  A tool for the C programming language, Tracer,  abstracts the code and uses a seq2seq model on the source code abstractions that are later concretized~\cite{ahmed2018compilation}.

All the DNN based Automatic Program Repair (APR) tasks have a fault localization step~\cite{tufano2019learning,chen2019sequencer,DLFix2020,ding2020patching,lutellier2020coconut}, and these tools' performance depends a lot on the fault-localizer. Semantic code correction is an inherently difficult problem, and syntax correction can be considered as a subset of semantic code correction problems. None of the previous syntax correction tools has compared their work with these tools because previous syntax correction tools did not depend on any fault localizer. Some of the APR tools~\cite{le2011genprog,DLFix2020, chakraborty2018tree2tree,long2016automatic} expects syntactically correct programs and those approaches are not applicable for syntactical code correction. For our purposes the most directly compatible recent APR tool was 
``SequenceR"~\cite{chen2019sequencer} which reported good performance,
and also fixes errors at the line level; it was readily adapted to using  the \javac to locate the line to be fixed, so we chose it
for comparison. Pradel et al. also detect specific types of bugs (e.g., accidentally swapped function arguments, incorrect binary operators, and incorrect operands in binary operations) but in syntactically correct code~\cite{pradel2018deepbugs}. 

Brown \etal used BlueJ IDE to collect the data in Blackbox repository~\cite{brown2014blackbox}
In this paper, we did a case study on the performance of the popular IDEs (e.g., Eclipse, IntelliJ, VSCode, BlueJ) in fixing novice programs. 
We compare repair hints from Eclipse JDT Core Compiler for Java (ECJ) (used in both Eclipse and VSCode) and javac (used by IntelliJ and BlueJ). That is, both Eclipse and VSCode present the same error messages, and IntelliJ and BlueJ present the same error messages. Four IDEs, but ultimately, only two compilers. \newname improves upon repair hints from both compilers.
Therefore, we primarily focus on Eclipse and IntelliJ for the case study. We chose VSCode because it is popular, well-documented,  available free for students, and is easy to extend. We were able to integrate \newname into VSCode without  any major difficulties.

\section{Conclusion}
We have described \newname, a machine-learning based tool to fix syntax errors in
programs. \newname leverages RoBERTa pre-training, uses compiler errors (both location and message),
and generates fixes using multi-label classification, rather than autoregressive generation,  to achieve
substantial improvements in fixing syntax errors. Our evaluation shows substantial improvements in fixing
rates over the previous best results reported by \oldname, and other tools, at all program lengths. Our evaluations suggest
that the the use of compilers to locate the precise line provides a big advantage; our evaluations also suggest
that the compiler error message \emph{per se} may be helpful in locating the precise token within the line
that needs to be repaired.  We have built
\newname into the VSCode IDE, and have found that even without a GPU, the \newname-enhanced
VSCode can fix syntax errors fairly quickly, often in less than a second. We have made all the source-code and data available, 
to the extent allowable under UK Law applicable to the  Blackbox dataset. \newname can fix errors that
IDEs (Eclipse, IntelliJ, BlueJ, and VSCode) cannot. In the supplementary materials (\textcolor{blue}{\href{https://bit.ly/3CMM0TP}{https://bit.ly/3CMM0TP}}) 
we show several real-world examples  of novice-made errors that cannot be fixed by any of these IDEs,
but can be fixed by \newname. 
Finally, the entire source for our \newname, including the VSCode extension,
is made available anonymously at \link{\url{https://doi.org/10.5281/zenodo.4563241}}. 

\ifCLASSOPTIONcompsoc
  \section*{Acknowledgments}
\else
  \section*{Acknowledgment}
\fi

This material is based upon work supported by the U.S. National Science Foundation under Grant Nos. 1414172, and 2107592. Any opinions, findings, and conclusions or recommendations expressed in this material are those of the author(s) and do not necessarily reflect the views of the National Science Foundation. Ahmed was also supported by UC Davis College of Engineering Dean’s Distinguished Fellowship.

\ifCLASSOPTIONcaptionsoff
  \newpage
\fi



\bibliographystyle{IEEEtran}
\bibliography{fromold.bib}
%



%

\begin{IEEEbiography}[{\includegraphics[width=1in,height=1.25in,clip,keepaspectratio]{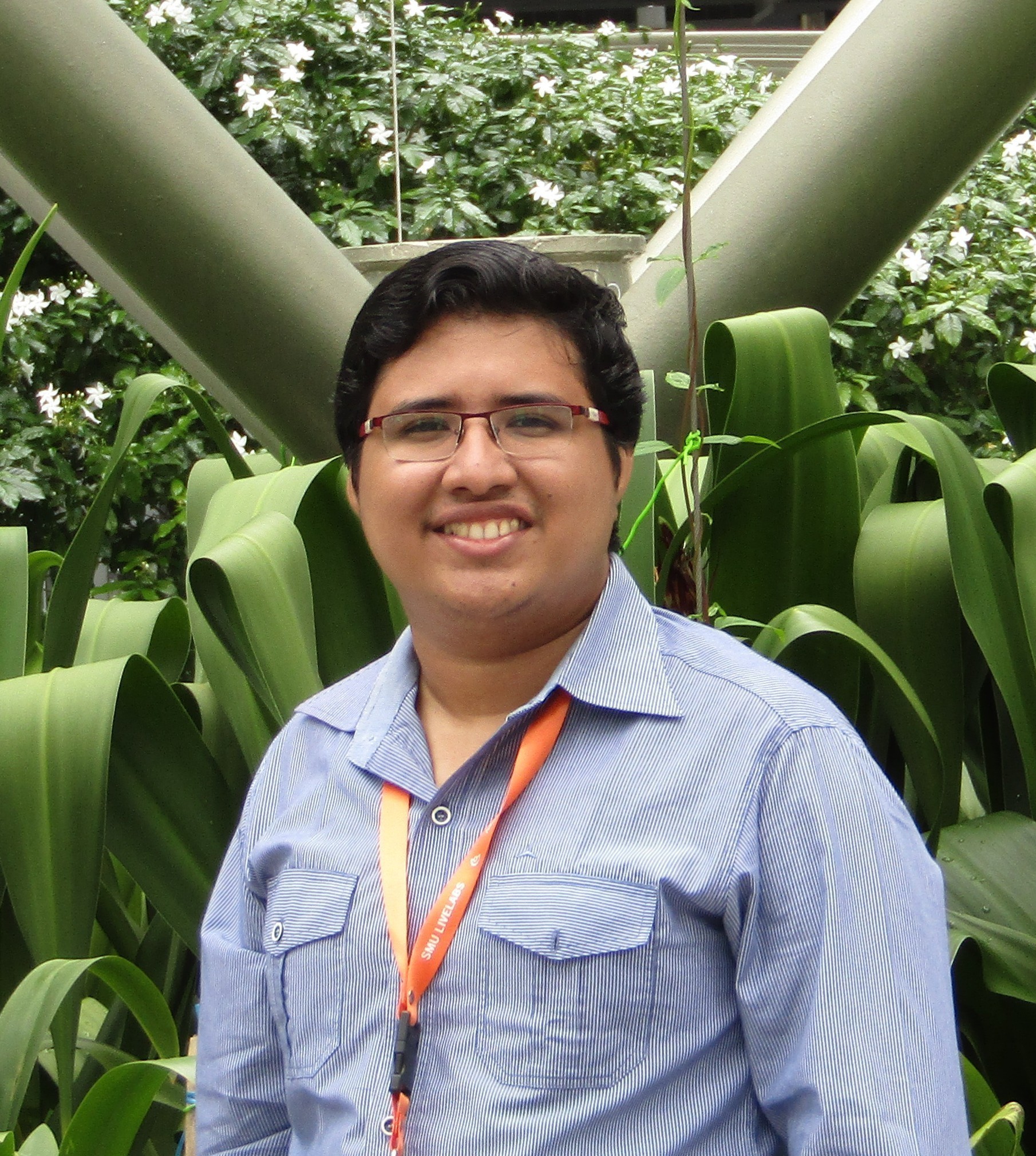}}]%
{Toufique Ahmed}
is a Ph.D. student at UC Davis. He received his B.Sc. and M.Sc. in Computer Science and Engineering from Bangladesh University of Engineering and Technology (BUET) in 2014 and 2016. He is currently working toward the PhD degree with University of California, Davis (UC Davis). His research interest includes Software Engineering, the Naturalness of Software, Machine Learning, and Sentiment Analysis. He is the recipient of the five-year prestigious Dean’s Distinguished Graduate Fellowship (DDGF) offered by The Office of graduate studies,  The College of Engineering, and The Graduate Group in Computer Science, UC Davis.

\end{IEEEbiography}

\begin{IEEEbiography}[{\includegraphics[width=1in,height=1.25in,clip,keepaspectratio]{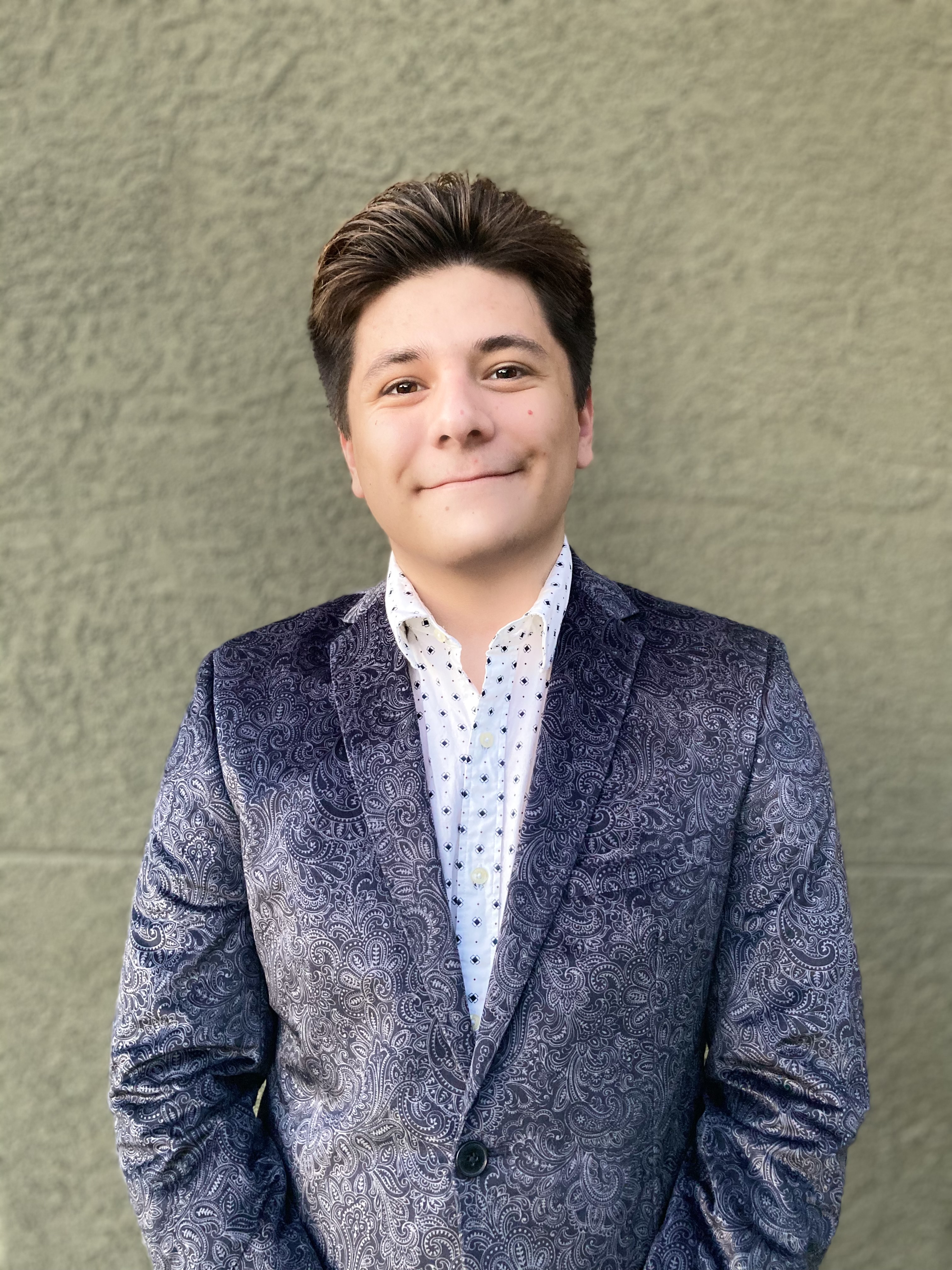}}]
  {Noah Rose Ledesma} is an undergraduate Computer Science student at the UC Davis, and a research assistant at  the DECAL lab. He is interested in machine learning, computer graphics, autonomous vehicles, and other subjects. Recently, Noah participated in an internship in which he developed infrastructure for Google Cloud's artificial intelligence platform. He will begin his career as a full-time software engineer after he graduates in June 2022.

\end{IEEEbiography}

\begin{IEEEbiography}[{\includegraphics[width=1in,height=1.00in,clip,keepaspectratio]{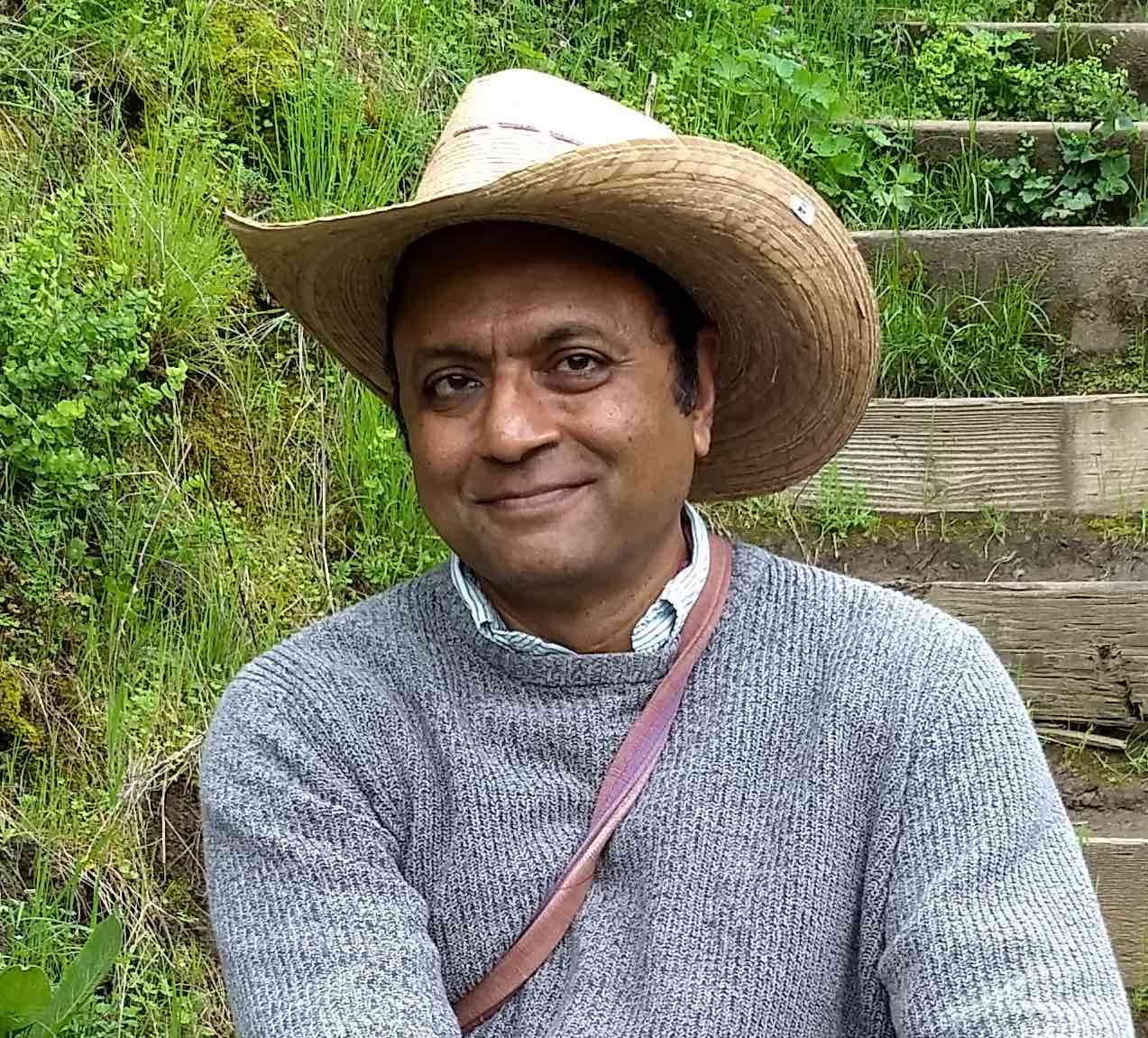}}]
{Premkumar Devanbu}
got his B. Tech from IIT Madras and his Ph.D from Rutgers University. He is currently Distinguished Professor of Computer Science at UC Davis. His research interests include Empirical Software Engineering and the applications of the Naturalness of Software.  He is an ACM Fellow. 

\end{IEEEbiography}




\end{document}